\documentclass[12pt,preprint]{aastex}

\shorttitle{{\it Chandra}'s tryst with SN 1995N}
\shortauthors{Chandra et al.}

\begin{document}

\title{{\it Chandra}'s tryst with SN 1995N}

\author{Poonam Chandra}
\affil{Tata Institute of Fundamental Research,
Mumbai 400 005, India }
\affil{Joint Astronomy Programme, Indian Institute of Science,
    Bangalore 560 012, India; {\tt poonam@tifr.res.in}}

\author{Alak Ray}
\affil{Tata Institute of Fundamental Research, Mumbai
400 005, India; {\tt akr@tifr.res.in}}

\author{Eric M. Schlegel}
\affil{Harvard-Smithsonian Center for Astrophysics, 
Cambridge, MA 02138, U.S.A.; 
{\tt eschlegel@cfa.harvard.edu}}

\author{Firoza K. Sutaria}
\affil{Technical University of Munich,  85748 Garching, Germany;    
{\tt fsutaria@ph.tum.de}}

\and

\author{Wolfgang Pietsch}
\affil{Max-Planck Institute for Extraterrestrial Physics, 
85748 Garching, Germany;
{\tt wnp@mpe.mpg.de}} 

\begin{abstract}

We present the spectroscopic and imaging analysis of a type IIn supernova 
SN 1995N observed with the Chandra X-ray observatory on 2004 March 27. 
 We compare the spectrum obtained from our Chandra observation with that
of the previous observation with ASCA in 1998.
We find the presence of Neon lines 
in the Chandra spectrum that were not reported in the ASCA observation. 
We see no evidence of Iron in both epochs.  The observed absorption column
depth  indicates an extra component over and above the
galactic absorption component and is possibly
due to a cool dense shell between the reverse-shock and the contact
discontinuity in the ejecta. The 
ASCA and the ROSAT 
observations suggested a non-linear behavior  of the X-ray light curve.
However, with the higher spatial 
resolution and sensitivity of Chandra, we  separate out many nearby 
sources in the supernova field-of-view that 
had additionally contributed to the 
supernova flux due to the large Point Spread Function of the ASCA. 
Taking out the contribution of those nearby 
sources, we find that
the light curves are consistent with a linear decline
profile. We consider the light curve in the high energy band separately.
We discuss 
our results in the context of models of nucleosynthesis and the
interaction of the shock waves with the circumstellar medium in
core collapse supernovae.

\end{abstract}


\section{Introduction}

X-rays from a supernova explosion arise from the interaction of the supersonic
ejecta with the circumstellar medium (CSM). The CSM
typically consists of a slow-moving
wind with the density $\rho=\dot{M}_{\odot}/4 \pi r^2 v_w$, 
where $ \dot{M}_{\odot}$ is
the mass loss rate, $r$ is the distance from the supernova and $v_{w}$ is the
wind velocity.
When the ejecta collides with the CSM,
it creates two shocks: a high-temperature, low-density, forward-shock
ploughing through the CSM (known as  blast-wave shock)
 and a low-temperature, high-density, reverse-shock
moving into the expanding ejecta. Initially the X-rays come
from the forward-shocked shell dominated by continuum radiation, but later on
 X-rays arise also from the reverse-shock, which can have 
substantial line emission, thus providing nucleosynthetic
fingerprints of the ejecta.
The temporal evolution of the X-ray luminosity of a supernova
can yield information on the density distribution
in the outer parts of the exploding star  ($\rho \propto r^{-n}$, here
$n$ can be in the range $7-12$, typically $n \sim 10$ for a Blue
Supergiant (BSG) and $n \sim 12$ for a Red Supergiant (RSG)- see
\citet{che03}). These studies  therefore are 
of interest from the stellar structure 
and the evolutionary point of view.

SN 1995N was discovered in MCG-02-38-017 (Arp 261) on 1995 May 5
\citep{pol95} at a distance of 24 Mpc (see \citet{fra02}).
\citet{pol95} estimated that the supernova was at least 10 months old upon
discovery, by comparing with spectroscopic chronometers. We
assume the date of explosion to be 1994 July 4 throughout the
paper. To date twenty-two supernovae have been detected in the
X-ray bands \footnote {See S. Immler's X-ray supernova page 
http://lheawww.gsfc.nasa.gov/users/immler/supernovae\_list.html
and \citet{imm03}.} 
and SN 1995N appears to be
at the high end of the X-ray luminosity \citep{fox00}, which makes it 
suitable for study in the X-ray wave bands even at late stages.
Additionally, SN 1995N is one of the 
only six supernovae of type IIn from which X-ray
emission has been observed; the others are 
SN 1978K \citep{sch04}, SN 1986J \citep{sch95},
SN 1988Z \citep{fab96}, SN 1998S \citep{poo02} and SN 2002hi \citep{poo03}.

\citet{nom96}
have suggested a continuum of merged stars in binary systems
with different common envelope
masses as possible progenitors of various types of supernovae, in
particular types IIL, IIb and IIn. 
The evolutionary path of a close
binary system depends upon the initial mass ratio $q_0$ ($= m_2/m_1$,
where $m_2$ is the less massive secondary star and $m_1$ is the
more massive primary star) and on their initial separation $a_0$.
If the initial mass ratio is $q_0 \leq 0.4$, the mass transfer
is highly non-conservative and leads to the formation of a common
envelope and the subsequent spiral-in of the secondary towards the
core of the primary. 
The structure of the CSM from the
spiral-in may be asymmetric, where the ejected mass may form a
bipolar jet or a disk-like structure. 
When the stellar core collapses and explodes, the supernova lights up 
the slow-moving gas into narrow emission lines leading to the type IIn
supernova classification (n for narrow emission line).
Type IIn supernovae 
show unusual optical characteristics and are known
to span a very broad range of photometric
properties such as decline rates at late times \citep{fil97}. 
It is likely that these differences are related to
their progenitor's structure, mass, composition as well as 
the composition and the density profile of the CSM \citep{li02}.
These supernovae show the presence of strong, narrow Balmer line
emission on top of the broader emission lines in their early spectra.
It is believed that the narrow emission lines originate
in the dense and ionized circumstellar (CS) gas \citep{hen87,fil91}.
The presence of strong H$\alpha$ emission line, the high bolometric
luminosity and the broad H$\alpha$ emission base 
powered by the interaction of the supernova shock with the CSM, all
point towards a very dense circumstellar environment \citep{chu03}. 
This interaction of the supernova shock with
the dense CS gas is indicated by strong radio and X-ray emission
detected from  several type IIn supernovae and in particular SN 1995N.
Optical and ultraviolet observations of SN 1995N \citep{fra02}
upto about 1800 days after the explosion, revealed three distinct
velocity components. 
Narrow lines from the CS gas show both low and high ionization
states and are caused by photo-ionization of the CS by X-rays from the
shock. The intermediate component has  a velocity of 
$2500-5000$ km s$^{-1}$, and is dominated by the newly processed
Oxygen that is conjectured by \citet{fra02}
to be a part of the unshocked (by the reverse-shock) ejecta.
The broad component whose extended wings reach $\sim 10,000$ km s$^{-1}$
is dominated by lines of HI, HeI, MgII, and FeII.

Apart from the narrow lines of H$\alpha$ mentioned above,
SN 1995N showed narrow lines of several other elements in multiple
charge states. In particular, the intensity ratios of
narrow Oxygen lines suggested a high electron density,
$n_e \geq 10^6 \rm cm^{-3}$ \citep{gar95}.
SN 1995N has turned out to be relatively bright in all the 
wavebands. It has been detected in radio wave bands with the VLA 
\citep{van96} and the Giant Metrewave Radio Telescope (GMRT) \citep{cha05}.
In X-ray wave band, it has been detected by  ROSAT in August 1996
and in August 1997 and later by  ASCA in January 1998 
\citep{fox00}.  The ROSAT observations  showed a 30\% decline
in the X-ray flux from August 1996 to August 1997, whereas the
ASCA observations showed a rise by a factor of two
in between August 1997 to January 1998 (see Table \ref{all_obs}).
To address the question of non-steady decline of 
the luminosity unequivocally, however, it is important to 
spatially resolve the field of view (FoV) around SN 1995N to an 
accuracy substantially larger than the ASCA angular resolution.
If this trend is indeed real, it could have very interesting implications
for the supernova.  It could imply that the X-ray emission may be coming 
from the clumpy clouds crushed by the forward-shock or it could 
show the inhomogeneity in the reverse-shock itself. Fig. \ref{models}
illustrates two different scenarios of the X-ray emission: one
due to the emission from the reverse-shock heated ejecta, which predicts 
linear decline in luminosity light curve,
and the other due to slow shocks in the CSM clumps, predicting bumps in the 
light curve. 

SN 1995N has declined very slowly in the optical band (see \citet{li02}
and references therein), with only a 2.5 mag change in the V-band over
$~ 2500$ days after explosion. This  is consistent with the
slow spectral evolution reported in \citet{fra02}.
Ground based optical and HST observations of the late-time spectral
evolution of SN 1995N were used by \citet{fra02} to argue
that the late-time evolution is most likely powered by the X-rays
from the interaction of the ejecta and the circumstellar medium
of the progenitor. They in turn proposed that the progenitors
of type IIn supernovae are similar to red supergiants in their
superwind phases when most of their hydrogen-rich gas is expelled
in the last $10^4\, \rm yr$ before explosion.

We observed SN 1995N under Cycle-5 Chandra Guest Observation program
for 55.74 ks
exposure time on 2004 March 27  with ACIS-S3. In this paper we
present the analysis and interpretation of the data and their
implications. 
Section 2 deals with the observation and data analysis. In Section 3, 
we present the results of spectroscopic and imaging analysis.
Section 4 gives a discussion of the results and  Section 5
presents a summary.

\section{Observations and data analysis}

We observed SN 1995N  from 2004 March 27 UT 17:55 to 2004 March 28 UT 10:00
with the Advanced CCD Imaging Spectrometer (ACIS-S) instrument of
Chandra X-ray observatory as a part of GO observations 
(Obs ID 5191). The CCDs S2, S3, I2 and I3 were switched on for the
observations with aim point on the back-illuminated chip S3.
The back-illuminated chip has an advantage 
of relatively flat spectral resolution
over the front-illuminated chips. The observation was taken in 
very faint (VF) mode. The total exposure on the source 
was 55.74 ks. The summary of the Chandra observations along with 
all previous X-ray observations of SN 1995N are provided in Table 
\ref{all_obs}. 

The data were analyzed using CIAO analysis threads and XSPEC \citep{arn96}. 
Event-2 files (pipeline processed files) were used for 
the data analysis. In order to check for any X-ray flaring events, 
we selected a background region of
size 87$''$, carefully excluding point sources. 
We then obtained the light curve of this selected background region
using ``{\it dmextract}" and CIAO-script ``{\it analyze\_ltcrv.sl}". 
We did not find any 
flaring and the data were quite clean and contained
   no spurious background events.

We then extracted the SN1995N events in a radius of 7$''$ centered at 
SN1995N's position. The total counts for the source events was 758 and the 
observed count rate 
was $0.0138\pm0.0026$ cts/s. We also extracted an annulus of
inner radius of 7$''$ and outer radius of 22$''$ centered at the
supernova position
to obtain the  background counts. 
We then generated the ``pulse invariant" spectra for 
the source and the background events using ``{\it dmextract}".
The response matrices were constructed at the position of SN 1995N. 
 We corrected the 
response matrix in order to take into account the contamination due to 
the ACIS quantum efficiency degradation at low energies.  To
generate the response matrices (RMF and ARF), we used the 
CIAO threads: 
``{\it acis\_fef\_lookup}", ``{\it mkrmf}", ``{\it asphist}", and 
``{\it mkarf}".

\subsection{Data Analysis: Spectral lines}

We binned the data in 20 cts per channel for the spectral analysis
using the HEASARC tool ``{\it GRPPHA}".
We used the XSPEC software  for the spectral fitting of the data. 
We ignored the counts below 0.3 keV and above 7.5 keV.
This is because there was 
almost no flux in this energy range and the Chandra response is 
quite poor at these energies.  
We modeled the background with a  best-fit
broken powerlaw model with a break at an energy 2.7 keV, and remove
its contribution from the source spectrum.
Table \ref{bkd1} gives the best-fit parameters for the background model. 

\subsection{Imaging Analysis}

We imaged the field of view of the SN 1995N.
We used only the S3 chip for imaging, the aim point for SN 1995N. 
We did not apply the CTI (Charge Transfer Inefficiency) correction 
to the data because S3-back illuminated chip is not significantly  
affected by this effect. 
Also we did not have enough counts to be able to see this tiny correction
factor. We built the relevant instrument map for the data using
{\it mkinstmap} and then
generated the spectrally weighted exposure map using {\it mkexpmap}. 
This file was generated using the 
CIAO-script {\it ``spectrum.sl"}. The event-2 image divided by the
exposure map gives the corrected image. We also ran ``{\it wavdetect}" to 
detect the faint sources in the field of view. 

\section{Results: X-ray spectra and High Resolution Image}
\subsection{Spectral fits to the data: line emission}

While fitting the source data models in our Chandra spectrum, 
we initially fixed the background to the 
values derived in Section 2.1 (Table \ref{bkd1}), and later, 
while doing the final iteration, 
we allowed these parameters to vary  to get the real estimates
on the errors of the source model parameters.
First we fit the continuum thermal 
bremsstrahlung model including the galactic absorption. 
This fit is certainly not a good fit. 
The spectral shape suggests the possibility
of a line around 1 keV. We then add a Gaussian component at 
1 keV and refit the 
data. The fit improves with the total ${\chi}^2$ changing
by 18, so the line parameters have a 
confidence level of 99.9\%. The best-fit line energy is 1.02 keV. 
We fit the Gaussian with its line width fixed at
zero.  When allowed to vary, the line width 
remains close to zero within the errors because the ACIS resolution
is insufficient to resolve the line. 
We further note the possibility of another line around 0.9 keV.
Since both the lines are very closely spaced in energy, it is
difficult to fit both the lines together.
Therefore, fixing the parameters of the Gaussian already fit at 1.02 keV, 
we add another Gaussian at 0.9 keV, again with the line 
width of the Gaussian fixed to zero. The $\chi^2$ improves by 5, 
significant at 90\% level for 2 parameters of 
interest (line position, normalization). The best-fit line 
energy is 0.85 keV.
Although this is not a very robust detection, we suggest the possible 
identification of this line as Ne~IX. Alternatively,
it could be a signature of the Ne~I K-edge. 
In Table \ref{bestfit_models}, we report the best-fit values 
for the various models. 
In Table \ref{bestfit_models} the best fit $\chi^2$ is obtained
for the reported parameters of 
Bremsstrahlung model ($kT=2.35\,{\rm keV}$, 
$N_H = 1.51 \times 10^{21}$ cm$^{-2}$)
with only 2 Gaussian components. The data and our model
are shown in Fig. \ref{bremsstrahlung}.
The corresponding confidence contours of the line  at 1.02 
identified as NeX K-$\alpha$ line are shown in 
Fig. \ref{fig:bremss+gaus1} 
and this line is very well constrained in the line centroid and 
normalization.
Fig. \ref{fig:bremss+gaus+gaus1} shows the confidence contours for the
line at 0.85 keV,
demonstrating that the line is reasonably  constrained 
at the 90\% confidence level in the centroid and normalization. 
We also plot the confidence contours of 
galactic absorption versus the bremsstrahlung
temperature. Both the parameters are well constrained
(see Fig. \ref{nh_kt}).

We tried to fit other lines  as well but we obtained only the 
upper limits, and our 
fits did not improve significantly via a  change in $\chi^2$. 
Table \ref{other_lines} gives the upper limits on O, Mg, Si, S and Fe lines.
Only Si line at 2 keV shows little significance; 
however, when we plot the confidence contours
for Si line, the lower bound of the line is completely unrestricted
even at 90\% level.
Fig. \ref{Si} shows the confidence contours for Si line, which clearly rules 
out the possibility of presence of the Si feature at 2 keV.

We also applied the VMekal model  to the spectrum. `VMekal' is a model in 
the XSPEC for an optically-thin
thermal plasma model developed using codes of \citet{kaa92,mew85,mew86,lie95}. 
The `L' in `Mekal' refers to the iron L shell
corrections to the codes from \citet{lie95}. We fit a 
model to the spectrum with a fixed hydrogen density of
$2 \times 10^6 \, {\rm cm^{-3}}$, with a solar abundance.
We find that the VMekal model fits are least sensitive to the 
hydrogen density. Here the
abundances of C, N, Na, Al, Ar, Ca and Ni  were fixed to the solar
abundance.
We separately let the abundances of  
O, Mg, Ne, Si, S and Fe vary. 
Only the abundance of Ne and Si are significant with respect to
solar (see Table \ref{bestfit_models} and 
\ref{other_lines}).  The enhanced Ne
is consistent with the results obtained from the bremsstrahlung plus 
Gaussian fit, so we identify the 1.02 keV emission as Ne X.
The presence of the Si line is ruled out from the confidence 
contour plot as discussed. Table \ref{other_lines}
show the upper limits on the abundances of various lines.
 The best fit column depth and temperature in the VMekal fit are:
$N_H=(3.71\pm2.79)\times 10^{20}\, {\rm cm^{-2}} $ and $kT=3.03$ keV.

Based on $\chi^2$ arguments, the bremsstrahlung plus two Gaussian model
provides a better fit to the data than VMekal model fit.
Table \ref{bestfit_models} provides the best-fit parameters for the VMekal. 
Fluxes predicted by various models  extrapolated to the ROSAT (0.1-2.4 keV),   
and ASCA (0.5-7.0 keV) bands are given in Table \ref{fluxes}.
 The fluxes quoted here are absorbed fluxes.
It is noticeable that in both the models, the fluxes predicted are 
quite similar to each other. 

The value of best fit absorption column density  in Bremsstrahlung model is
$N_H = 1.51 \times 10^{21} \, {\rm cm^{-2}} $. 
From the galactic extinction maps, 
we get $E_{B-V}=0.1158$ for SN 1995N  coordinates, 
which gives galactic absorption 
$N_H = 5.3 \times 10^{21} E_{B-V}$, using the relation of \citet{pre95},
to  $6.1 \times 10^{20} \, {\rm cm^{-2}}$. The latter is about two and a 
half times smaller
than the value we obtain from the model fits. We tried to fit the above 
models again by fixing the galactic absorption to be 
$6.1 \times 10^{20}\, {\rm cm^{-2}}$. Our results were much worse than 
the previous fits. 
Table \ref{bestfit_models} gives best-fit parameters for the model fit
when $N_H$ is fixed to $6.1\times 10^{20} {\rm cm^{-2}}$.
Here we mention that the $N_H$
determined from radio measurements of the galactic 21-cm line \citep{dic90}
is $7.8\times 10^{20} {\rm cm^{-2}}$, which is consistent with the
$N_H$ determined from optical extinction (see $N_H$ from HEASARC page). 
 We considered the possibility that the extra column depth in
our best fit models is due to the
interstellar medium of the host galaxy. However, the Bremsstrahlung 
or powerlaw models fitted to sources other than SN 1995N
in the field of view of ACIS yield a column 
depth of $N_H = 6.97\times 10^{20} \; {\rm cm^{-2}}$ (see Section 3.2), which
is same as that obtained from galactic extinction calculations. 
This implies that
there is almost no extra absorption due to the interstellar medium of the host
galaxy. Therefore the excess $N_H$ seen in the Chandra spectrum of SN 1995N 
over the galactic and host galaxy column depth is most likely
in the SN itself, due to the cool
ejecta shell formed after the passage of the reverse-shock through it
inwards of the contact discontinuity. 
 If the extra absorption layer ($N_H = 1.5 \times 10^{21} \; \rm cm^{-2}
- 6 \times 10^{20} \; \rm cm^{-2} = 9 \times 10^{20} \; \rm cm^{-2}$)
were located deep in the He layers
near the external boundary of the C+O core, with a
characteristic expansion velocity of $5000\, \rm km s^{-1}$ (seen in
the Oxygen rich layers in the Optical/UV bands
by \citet{fra02}); 
then the upper limit to the
mass swept-up in the cool shell is (\citet{che03}): 

$$M_{rev}=N_{cool}4\pi R_s^2 m_p (X_O^{\odot}/X_O^{sh})=
2 \times 10^{-3}M_{\odot}$$
since it is the Oxygen abundance of the gas ($X_O^{sh}/X_O^{\odot}=
100$) that mainly contributes to the
X-ray absorption near 1 keV.  If the cool shell is well into the helium
rich layers, then the implied mass in the absorption shell is considerably
larger ($\sim 0.8 M_{\odot}$). 

\subsubsection{Density and ionization state of the radiating plasma}

An important question in studying the radiative properties of an ionized 
plasma is the mechanism of ionization, which we shall discuss elsewhere 
(paper in preparation). 
The strengths of the Ne~X line reported in 
section 3.1 (see Table \ref{bestfit_models}) allow us
to deduce whether the plasma is in 
an equilibrium ionization state.
The relaxation time for a particular ion, $t_{rel}$ is 
minimum of the ionization time ($t_{ion}=[C(T_e)n_e]^{-1}$) \citep{mew99} and
recombination time ($t_{rec}=[\alpha(T_e)n_e]^{-1}$), where 
$C(T_e)$ and $\alpha(T_e)$ are the collisional ionization
and recombination coefficients, and $T_e$ is the electron temperature. 
The re-ionization parameter
is given by \citep{ver96}
\begin{equation}
\alpha(T_e)=a[(T_e/T_0)^{1/2}(1+(T_e/T_0)^{1/2})^{1-b}(1+(T_e/T_1)^{1/2})^{1+b}]^{-1}
\end{equation}
where $a$, $b$, $T_0$ and $T_1$ are best-fit parameters, 
obtained from \citet{ver96} for Ne~X, and leads to a recombination 
parameter $\alpha(T_e)=2.8\times10^{-13}\,{\rm cm^3\,s^{-1}}$.
Hence $n_et_{rel}={\rm min}(1/C(T_e), 1/\alpha(T_e))=
1.32\times10^{12} \,{\rm cm^{-3}\,s}$. 

For a steady state ionization to hold, the relaxation time for ionization
should be shorter than the local cooling and the 
expansion times. 
The cooling timescale $t_{cool}$, given by:
$n_e t_{cool}= 3kT/\Lambda$, where
$\Lambda$ is cooling function, can be obtained from
\citet{mcc87} :
\begin{equation}
\Lambda=1.99 \times 10^{-23} T_7^{-0.7} + 7.27 \times 10^{-25} T_7^{-1/2}\,
{\rm ergs\, cm^2\,s^{-1}}
\end{equation}
where $T_7$ is the temperature in $10^7$ K. Hence we get 
$\Lambda=1.03 \times 10^{-23} \, {\rm ergs\, cm^2\,s^{-1}}$.
Therefore $n_e t_{cool}=1.1 \times 10^{15} \,{\rm cm^{-3}\,s}$,
approximately the same as the corresponding 
hydrodynamic timescale parameter and
which is much larger than the relaxation time. Hence the plasma is
in ionization equilibrium, which is expected for a high-density young 
supernova. 
We note that the above  analytical expression for cooling does not
hold for non solar composition. However, Nymark, Fransson and
Kozma (private communication) have shown that the equilibrium is
nevertheless expected, except for low temperature shocks and shocks
in medium highly enriched in heavy elements (e.g. oxygen) and 
equilibrium does not 
hold good for lines with $h \nu \leq 0.1$ keV.

\subsubsection{Line luminosity, Neon mass and the location of the emitting gas}

The line luminosity for a given line can be given as volume integral of
its emissivity. For NeX line:
\begin{equation}
\label{emissivity}
L_{NeX}=\eta \int\int j_{NeX} {\rm d}\Omega {\rm d} V
\end{equation}
where $\eta$ is a cascade factor that denotes the fraction of
recombinations that leads to emission of the Lyman $\alpha$
line in Neon. $\eta$ is at least 0.13 (the number of direct
recombinations to ($n=2 \, ^2P$) excited state  is typically 13\% of
total, -- see \citet{ost89} Table 2.1) 
and ranges up to about $\sim 50\%$ for all final states except the
ground state (we assume $\eta=0.4$).
From Table \ref{bestfit_models}, the line strength of Ne~X 
($3.4\times 10^{-6}\, {\rm photons \, cm^{-2}\, s^{-1}}=
5.6\times 10^{-15}\,{\rm ergs\, cm^{-2}\, s^{-1}}$) leads to 
$L_{NeX}=3.9 \times 10^{38}\, {\rm ergs\,s^{-1}}$ at SN 1995N 
distance (24 Mpc).  Emissivity for Ne~X can be given as
\begin{equation}
j_{NeX}=n_e n_{NeXI} \alpha_{NeX}^{eff} \frac{h \nu_{NeX}}{4 \pi}
\end{equation}  
Let us assume $n_{Ne~XI}$ is a fraction $f$ of total $n_{Ne}$
i.e. $n_{Ne~XI} = f \; n_{Ne}$.
If most of the Ne is in the Helium zone close to the C+O core boundary, 
(see the Discussion section) then with $n_{Ne}=6.65\times 10^{-4} \, n_{He}$
(see \citet{woo02}, Fig. 9 for a 15 $M_\odot$ star) and $n_{e} = 2 n_{He}$,
the emissivity of Ne~X (see the above two equations) gives: 
\begin{equation}
\eta \int n_e^2 {\rm d}V= \eta \frac{4 \pi R^3}{3} n_e^2 =
\frac{6.41 \times 10^{63}}{f} \, \rm cm^{-3}
\end{equation}
The first equality assumes that the gas is uniformly distributed.
Since the velocity at Oxygen layer is 5000 km/s, the velocity of the 
Neon has to be greater than this if the Neon is in the Helium layers. 
Hence for the present epoch, we assume $R=vt=1.57\times10^{17}(v/5000\,{\rm km\,s^{-1}
})\,(t/10\,{\rm yrs})$.
Therefore
\begin{equation}
n_e=\frac{6.77\times10^{5}}{\sqrt f} \, {\rm cm^{-3}}
\end{equation}
 The ionization potentials of all ionized Neon 
species upto Ne~VIII are less than  240 eV, whereas those of
Ne~IX and Ne~X are above 1195 eV. Hence at such high temperatures
as found by our Chandra observations, the predominant 
species of Neon are expected to be Ne~X and Ne~XI.
This is also indicated by the clear detection of spectral lines 
of Ne~X and  a possible detection of Ne~IX.
Thus we expect that the fraction of Ne~XI will be significant, i.e.
 $f=n_{Ne~XI}/n_{Ne}\ge0.1$. For $f=0.1$,
$n_e=2.1 \times 10^6 \, {\rm cm^{-3}}$, and 
$n_{He}= n_e/2=1.0 \times 10^6 \, {\rm cm^{-3}}$, which gives 
$n_{Ne}=687\, {\rm cm^{-3}}$. 
The corresponding mass of Ne, measured from the above Ne density,
 is $0.16 (f/0.1)^{-1/2} M_{\odot}$. 
Note that the implied number density of the electron is not so
 sensitive a function
of the ionization fraction of $Ne_{XI}$.

Table \ref{woosley}  gives theoretical estimates by \citet{woo02} of elements
synthesized in inner zones of massive stars which can be compared with
our estimate of the Neon mass detected in SN 1995N. 
However, the Neon mass estimated in the previous paragraph is 
an overestimate due to two reasons: firstly, the Neon is confined 
mainly in a thin layer, with a shell of  thickness $\Delta r$, and
not in the entire total volume inside the reverse shock;
secondly, we have assumed a constant density in the shell, whereas, the
density is a function of radius (velocity) as: $\rho \propto (1/t^3)r^{-n}$
 ($n\le8$ for adiabatic shock). 
The Neon layer lies between Oxygen (velocity $\sim 5000\,{\rm  
km s^{-1}}$) 
and Hydrogen-Helium zone (velocity $\sim 10,000\,{\rm  km s^{-1}}$), i.e. 
in a mass shell of
 3.05 $M_\odot-$4.1  $M_\odot$ for a 15 $M_\odot$ star (see Table
 \ref{woosley}). 
Hence, the Neon mass depends  on the velocity at which it occurs as: 
$$\int \rho dV \propto \int (1/t^3)r^{-n} \, 4 \pi r^2 dr=K\int^{v_1}_{v_2} 
v^{-n+2}dv$$
This velocity dependent factor 
reduces  by $\approx$ 30 from the Oxygen layer to the 
Hydrogen-Helium layer boundary for $n=8$. Hence the
actual mass of Neon is likely to
 be at least an order of magnitude lower than 
$0.16  M_{\odot}$ estimated 
above or about $5\times 10^{-3}M_{\odot}-
1\times10^{-2}M_{\odot}$. This is consistent with Neon being 
in the Helium layer for a $15 M_{\odot}$ star (see row 3 of
 Table \ref{woosley}). 
We find that for other stellar masses and other composition zones, the
required Neon mass is much larger than
observed. Therefore, Neon in the Helium core
of the $15 M_{\odot}$ star is the probable site where it was co-synthesized
with C, N, and He.

\subsection{Imaging the SN 1995N Field of View}

Fig. \ref{fov} inset shows 
the part of the SN 1995N FoV with optical contours overlaid. 
The positions of SN1995N and the galactic center of
MCG-03-38-017 are also shown in Fig. \ref{fov}.
The nearest detected source from the supernova at RA= 
$14^{\mathrm{h}}49^{\mathrm{m}}29\fs09$, and Dec= 
$-10\arcdeg10\arcmin37.38\arcsec$ at a distance of $26\arcsec$, is 20 times 
less bright (count rate of 0.0007 cts/s) than the supernova.
Thus, the resolution and sensitivity achievable by the Chandra are 
crucial to the proper measurement of the SN 1995N flux in this
complex field.

We note that in the ASCA observations, 
the supernova counts were extracted from a circle of radius 4$\arcmin$ 
  centered on the supernova
position \citep{fox00}, excluding the counts in a circle of radius
1.5$\arcmin$   centered on source ``A" (shown as the
non-circled source at the extreme right edge in Fig \ref{fov}). Fig. 
\ref{fov} shows the Chandra ACIS S3 image with a circle
of radius 4$\arcmin$ centered on SN 1995N. It is evident 
from the figure that even after excluding source ``A", there are 
many other sources within the 4$\arcmin$ circle
that must have contributed towards the supernova flux, including the
parent galaxy MCG-02-38-017. ROSAT-HRI had sufficient resolution 
to detect these sources. But due to its high 
instrumental background, it could detect only two sources apart from the SN, 
in the 4$\arcmin$ circle as opposed to
10 sources in Chandra observations. The sources detected by ROSAT-HRI were
galactic center (0.0009 cts/s) and the source "A" (0.0025 cts/s). The 
corresponding SN counts in the Aug 1996 ROSAT observation were 0.009 cts/s.   
In our Chandra observations, we extracted the counts 
from all the sources in HPD of 4$\arcmin$  and 
fit the Bremsstrahlung and power-law models to the 
summed spectrum assuming all the sources in 4$\arcmin$ radius
 as a single source (excluding the source ``A"). 
Since the power-law model fits gives a better 
${\chi}^2$ (model parameters: 
$\gamma=2.14^{+0.10}_{-0.15}$, $N_H= 6.97^{+2.4}_{-3.3}
 \times 10^{20}\,{\rm
erg\,cm^{-2}\,s^{-1}}$), we adopt power-law model. The 
observed counts of the combined sources is 2328, 
whereas the total counts  for SN 1995N is
758. This means that if we were to include all these sources, we 
would have over-predicted the supernova flux by a 
factor of 3 at the Chandra epoch. 

\subsection{X-ray light curve of SN 1995N}

Since the supernova was observed with the ROSAT in 1996 and 1997 and with
the ASCA in 1998, we construct the light curves 
over an interval of 8 years.
Fig. \ref{new_lc} shows the light curves for the unabsorbed fluxes
at soft (ROSAT) energies
(0.1-2.4 keV) and at hard (ASCA) energies (0.5-7.0 keV).
The ROSAT HRI fluxes were 
determined by \citet{fox00} using the ASCA spectrum. 
 We re-determined the
ROSAT fluxes and their uncertainties (ranges) from the ASCA spectrum
(see Table \ref{all_obs}). 
Although the fluxes remain identical, the uncertainties increased by
a small amount, particularly for the high-energy bands.
To calculate the
unabsorbed flux, we fixed the galactic absorption to zero and 
calculated the fluxes in the above two bands. The
light curves show that the ASCA flux is well above all the other
flux measurements. Without the ASCA point and only with the ROSAT and
Chandra data points, a consistently 
declining trend in the flux is apparent. 
If the trend is truly linear,  then the 
 expected flux at the ASCA epoch should be $\sim 6 \times 10^{-13}\,{\rm
erg\,cm^{-2}\,s^{-1}}$ in the 
0.1-2.4 keV band and $\sim 8  \times 10^{-13}\,{\rm
erg\,cm^{-2}\,s^{-1}}$ in the 0.5-7.0 keV band, approximately a factor of
1.6 below the observed ASCA data points (Fig. \ref{new_lc}). 
However in the Section 3.2, we have shown that 
due to the large half-power diameter of the 
ASCA mirror system, 
the measured flux of SN 1995N in January 1998 measurement was subject to 
contamination from the other luminous X-ray sources contained within
the angular footprint of the ASCA. Since our Chandra image identified and 
measured the fluxes from these sources, 
we can evaluate the apparent non-linear behavior of the light curve.

We find from the Section 3.2 
that if we assume that the X-ray flux of the combined sources 
remained constant with time, then it explains most of 
the ASCA flux excess, from that of the predicted one from 
the linear extrapolation between ROSAT and Chandra points, 
shown in Fig. \ref{new_lc}.
The unabsorbed flux from the Chandra spectrum
in the 0.1-2.4 keV band in a circle of HPD
4$\arcmin$ (excluding source ``A"), is $ (2.94 \pm 0.34)  
\times 10^{-13}\,{\rm  erg\,cm^{-2}\,s^{-1}}$. Since we know that the
contribution of the supernova flux is $ (0.85 \pm 0.20)
\times 10^{-13}\,{\rm  erg\,cm^{-2}\,s^{-1}}$ in this band, the excess flux 
due to the rest of the sources {\it at present} is $ (2.09 \pm 0.39)
\times 10^{-13}\,{\rm  erg\,cm^{-2}\,s^{-1}}$. Similarly in the 0.5-7.0
keV energy range, the unabsorbed flux in the 4$\arcmin$ radius is
$ (3.15 \pm 0.36) \times 10^{-13}\,{\rm  erg\,cm^{-2}\,s^{-1}}$. Excluding the
supernova contribution in this higher energy band, i.e., $ (1.05 \pm 0.25)
  \times 10^{-13}\,
{\rm  erg\,cm^{-2}\,s^{-1}}$, the flux of the rest of the sources is
 $ (2.11 \pm 0.44) \times 10^{-13}\,{\rm  erg\,cm^{-2}\,s^{-1}}$ at 
present.
If we assume that the flux of all these sources remained constant from
the ASCA epoch to the present Chandra epoch, 
then we can subtract this contribution from 
that of the ASCA measurement of the supernova,
to estimate the actual  contribution due to the supernova only.
By this method the supernova flux at the ASCA epoch in 0.1-2.4 keV band is
$ (6.9 \pm 1.6) \times 10^{-13}\,{\rm  erg\,cm^{-2}\,s^{-1}}$ and in
the 0.5-7.0 keV band, it is $ (10.5 \pm 0.7) \times 10^{-13}\,{\rm  erg\,cm^{-2}
\,s^{-1}}$. Fig. \ref{new_lc} shows that in the 0.1-2.4 keV band,
the ASCA corrected flux falls on the light curve within the error bars
but the discrepancy is larger at 0.5-7.0 keV. Thus the soft band
light curve appear consistent with
a linear decline, including the ASCA January 1998 measurement. 
However there   appeared to be a possibility that there was an
extra flux in the hard X-ray band. 

We also plot the de-reddened  V-band 
optical light curve \citep{li02} and H$\alpha$ light curve \citep{fra02}
 along with the X-ray decline rate (see
Fig. \ref{new_lc}). Here, $A_V=3 E_{B-V}=0.3474$. 
H$\alpha$ light curve dominates over the V-band light curve until $\sim 1200$
days since explosion.  Both V-band and H$\alpha$ fluxes are much lower
than X-ray fluxes.

\subsection{Hard X-ray excess in the ASCA era?}

There could be a few possible scenarios to explain this excess of
the hard X-ray flux.  
\begin{enumerate}
\item It is possible that the 
supernova flux had really increased at the ASCA epoch, and the increase
was essentially in the hard X-rays. In that case, it is   
a physically interesting feature and could be due to the
dense clumps from the ejecta being hit by  the reverse-shock. Alternatively,
it could be due to an inhomogeneous CSM being hit 
by the blast-wave shock.

\item Another possible explanation could be that
our assumption about the sources 
within the 4$\arcmin$ FoV of the supernova being
constant in time is incorrect. It is quite possible that one
or more of these sources are time variable, such as 
X-ray binaries, and  most of their flux  fall
in the  hard X-ray band. 
We estimate, that to explain
the excess flux of $\sim 2 \times 10^{-13}\,{\rm  erg\,cm^{-2}\,s^{-1}}$ 
in the $2.4-7.0$  keV band (as described in last section) 
in January 1998  ASCA observation
over a linear decline
between 1996 and 2004, a typical
X-ray binary contributing roughly  1000 counts is required. This 
corresponds to a count rate increase of $\sim$0.02 counts s$^{-1}$
in the Bremsstrahlung model with $kT=3.32 \pm 0.36$ and $N_H
=(8.42 \pm 1.40) \times 10^{20}\, {\rm cm^{-2}}$. 
Hence, such variability in one X-ray binary can explain the excess ASCA
hard band flux.
A detailed analysis of the sources in the Chandra FoV, other than SN 1995N
is in progress and will be reported elsewhere (Sutaria et al., in preparation).
\item 
In this scenario, we note that for the 
ASCA analysis of the supernova spectrum, the circle of 4$\arcmin$ HPD
was chosen for the supernova counts and then the contribution of 
source ``A" was corrected by excluding the counts from the
circle of 1.5$\arcmin$ HPD. We estimate that in view of 
ASCA's large PSF, 
a 1.5$\arcmin$ HPD circle around source ``A"
would have included only 40\% of hard photons coming from source 
``A". The remaining $\sim$60\% of the
hard photons would have appeared in the 4$\arcmin$ extraction circle and
would have contributed towards the supernova flux in 2-6 keV band.
Since source ``A" was very bright, the scattered counts 
could cause the high flux of the supernova in the
harder energy band and hence could explain the 
discrepancy of the corrected ASCA flux not falling on the linear
light curve in the 
harder X-ray bands. We estimated that the flux for the source ``A" in the 
Chandra observation, in 0.5-7.0 keV in Bremsstrahlung model 
was $1.8 \times 10^{-13}\,{\rm erg\, cm^{-2}\,s^{-1}}$, and 60\% of this
was $1.1 \times 10^{-13}\,{\rm erg\, cm^{-2}\,s^{-1}}$ which accounted
for a large fraction of the discrepancy. However,
we cannot accurately measure  the flux of 
source ``A" in our Chandra observations
because it is on the edge of the chip so our
estimate of the flux of source ``A" and consequently the correction on 
the ASCA
measurement would be an underestimate of the actual flux.
We found that the source ``A" had $\sim700$ counts in the Chandra data with 
most of the photons in the hard X-ray band;
the hardness ratio of the source was 
$\sim$+0.25 (using $(H-S)/(H+S)$, where $H$
were the counts in the 1.0-7.0 keV and $S$ the counts in the 0.5-1.0 keV band).
\end{enumerate}

\subsection{Comparison with ASCA spectrum: continuum and line fluxes}

SN 1995N was observed in 1998 with ASCA \citep{fox00}, although
that paper did not include any spectral fits containing line emission.
We compared the Chandra bremsstrahlung 
model with a fit to the ASCA SIS-0 spectrum
(SIS-0 provides the best spectral resolution) with the results
shown in the  upper panel of Fig \ref{fig:asca_bremss} and discussed below.
We used the screened event list available from the HEASARC database
and extracted the spectrum as described in \citet{fox00}.
  We applied the
best-fit Chandra 
 (pure bremsstrahlung)
spectrum 
to the
 ASCA data and the difference was visible in the
slightly harder ASCA spectrum, a line feature at  $1.0$ kev and at 
$\sim 1.3$ keV,
as well as the larger ASCA flux (Fig \ref{fig:asca_bremss}).  This
Chandra model yields a very poor fit to the ASCA data. We then 
fit the  
ASCA spectrum with bremsstrahlung model with N$_{\rm H}=(1.31\pm0.23 ) 
\times 10^{21}\,{\rm cm^{-2}}$ and
a temperature (${\rm kT}=
2.53\pm0.63 $ keV)  plus the two lines at 1.02 keV and 1.32 keV, 
which yields a drop in ${\chi}^2/{\nu}$ of
19.6, significant at more than 99.99\% for the 2 extra degrees of freedom
(lower panel of Fig \ref{fig:asca_bremss}).
Furthermore, the 0.85 keV line is consistent with being absent 
at 90\% with an
upper limit on the equivalent width of $\sim$40 eV.  We also detect a
line at 1.36 keV  at about 1 $\sigma$ level
with an equivalent width of 140 eV (90\% upper limit
of 190 eV) that is not detected in the Chandra spectrum (90\% upper limit
of 29 eV).   (The ASCA lines mentioned here had norms:
$5\times 10^{-5} (1.36 \;\rm keV)$ and $2.6 \times 10^{-4} (1.02 \;\rm keV)$).
We may interpret these results as an indication of the degree
of difference between the ASCA and Chandra spectra (the Chandra ACIS 
spectrum is displayed in Fig. \ref{bremsstrahlung} ).
However since the ASCA spectrum has contributions from sources other than
SN 1995N (see e.g. section 3.2 above), the small difference between
the ASCA and Chandra spectra may not be due the supernova itself. 

  The 1.02 keV line is detected in the ASCA data with an equivalent
width of $\sim$ 250$^{+600}_{-115}$ eV (compared to the
 $\sim$129$^{+83}_{-20}$ eV for the Chandra spectrum). If we attribute
the 1.02 keV line to Ne~ X and the 0.85 keV line to Ne~ IX, then the change
in line strengths may imply a decrease in the ionization conditions in
the ejecta.  However as pointed out above, within the errors the
values are identical (the error bars are just barely separated at 90\%),
or that line emission from X-ray binaries  within the ASCA PSF
contaminates the ASCA spectrum.

\section{Discussion}

\subsection{Spectrum, light curve and the site of X-ray line emission}

 The
Chandra spectrum differs from that of ASCA in slightly harder X-ray band,
with a larger ASCA flux and a line at $\sim 1.3$ keV, not seen in Chandra
spectrum.
The presence of at least one line or very likely
two lines of Ne in the spectrum indicates that the emitting gas has become
optically thin by now. 
We have a robust detection of one line at 1.02 keV and probable
detection of another line at 0.85 keV. At 1.02 keV, it could be Ne X or some 
of the higher ionized states of Iron. 
The 0.82 keV region contains lines of Fe~XVII to Fe~XX
\citep{lie92} that are expected to be strong at high temperatures and
low densities.  Detailed line calculations and spectra with higher
resolution are required to understand the exact range of possibilities.
However, the strongest Fe~XVII
feature in 0.82 keV does not show up in our spectrum, nor is there any 
evidence of Fe lines around 6.7 keV (see Table \ref{other_lines}). 
Absence of the 
very strong Fe~XVII line feature at 0.82 keV 
indicates that Iron is most likely absent in the spectrum  (i.e. unmixed
with lighter elements in the ejecta) or in a cold state.
Therefore, most likely, we are seeing the 
Ne~X (1.02 keV) and Ne~IX (0.9 keV) lines.
But along with the Ne lines, one would have expected to see the
 Oxygen lines as well around 0.6-0.7 keV.
However we do not see this in the Chandra spectrum. This
could be due either to the high 
galactic absorption in the low-energy bands and/or the lower sensitivity
of the ACIS detector below 0.7 keV.

We note that due to low counting statistics, we are unable to resolve
the line-widths of the Ne lines and thus cannot say if this gas is indeed
at a high ($\sim 10,000\, {\rm km \, s^{-1}}$) or  intermediate
velocity ($2500-5000\, {\rm km \, s^{-1}}$). If the Ne arises in the
O-Ne-Mg core or partially burnt C-shell and is shocked by
the reverse-shock then a high temperature  needed for the high
ionization state for X-ray emission as well as the broad emission 
line-widths seen in the optical-UV spectra \citep{fra02} 
are possible. Alternatively, it could be coming  from a shell of partially
burnt Helium in a shell burning
that is photo-ionized by the shock.  
Although SN 1995N is believed
to have lost most of its hydrogen rich envelope before the explosion,
and hence the relatively high velocity ($\sim 5000 \rm km \; s^{-1}$)
of the Oxygen core component in an ``untamped" explosion,
the UV optical spectrum does reveal
a high velocity ($v \sim 10,000 \rm km \; s^{-1}$) and high density
($n \sim 10^9 \rm cm^{-3}$)
Hydrogen-Helium dominated gas at low-ionization.

In our best-fits of spectrum, we find that the
absorption column density is at least
2.5 times more than that calculated from the galactic extinction maps. 
 The best 
fit model towards other sources do not show this extra absorption component.
This suggests that the moderate, extra absorption is likely to be 
due to the formation of a thin cool ejecta-shell after the reverse-shock.

 The light curves of
SN 1995N suggested a non-linear profile due to high ASCA flux. 
If the contributing factor for this jump (or shoulder) is the
supernova itself, it could have 
interesting implications for the CSM. 
We therefore re-analyzed the 
ASCA results in view of the high-resolution imaging data obtained
by Chandra. We find that
due to  ASCA's large PSF, at least ten more sources were
contributing to what was taken to be 
the supernova flux. 
 The analysis discussed in Section 3.3 and 3.4 indicates that the
luminosity light curve is consistent with linear 
decline within error bars in both hard and soft energy bands.

\begin{figure}
\centering
\includegraphics[angle=0,scale=0.75]{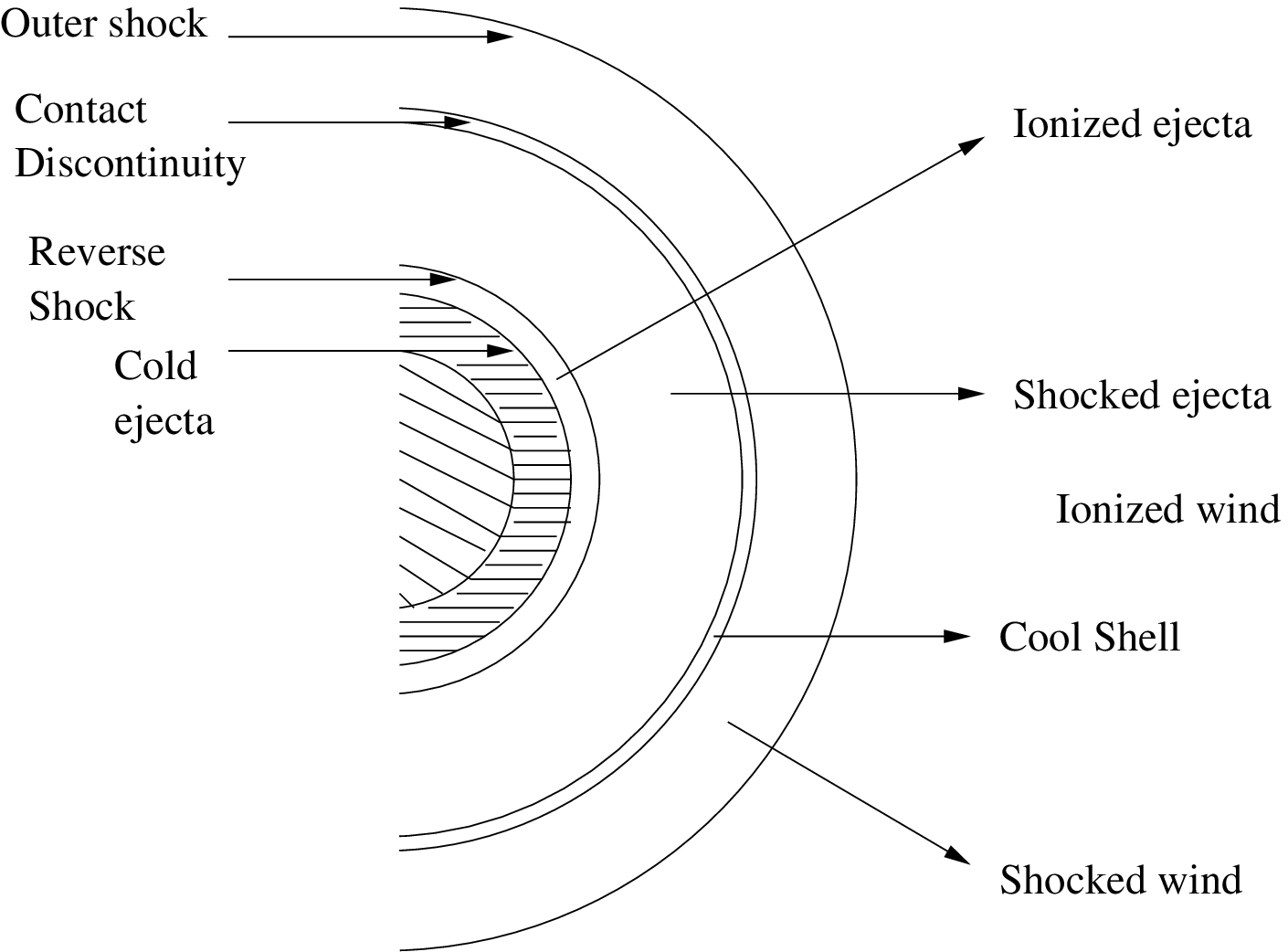}
\includegraphics[angle=0,scale=0.75]{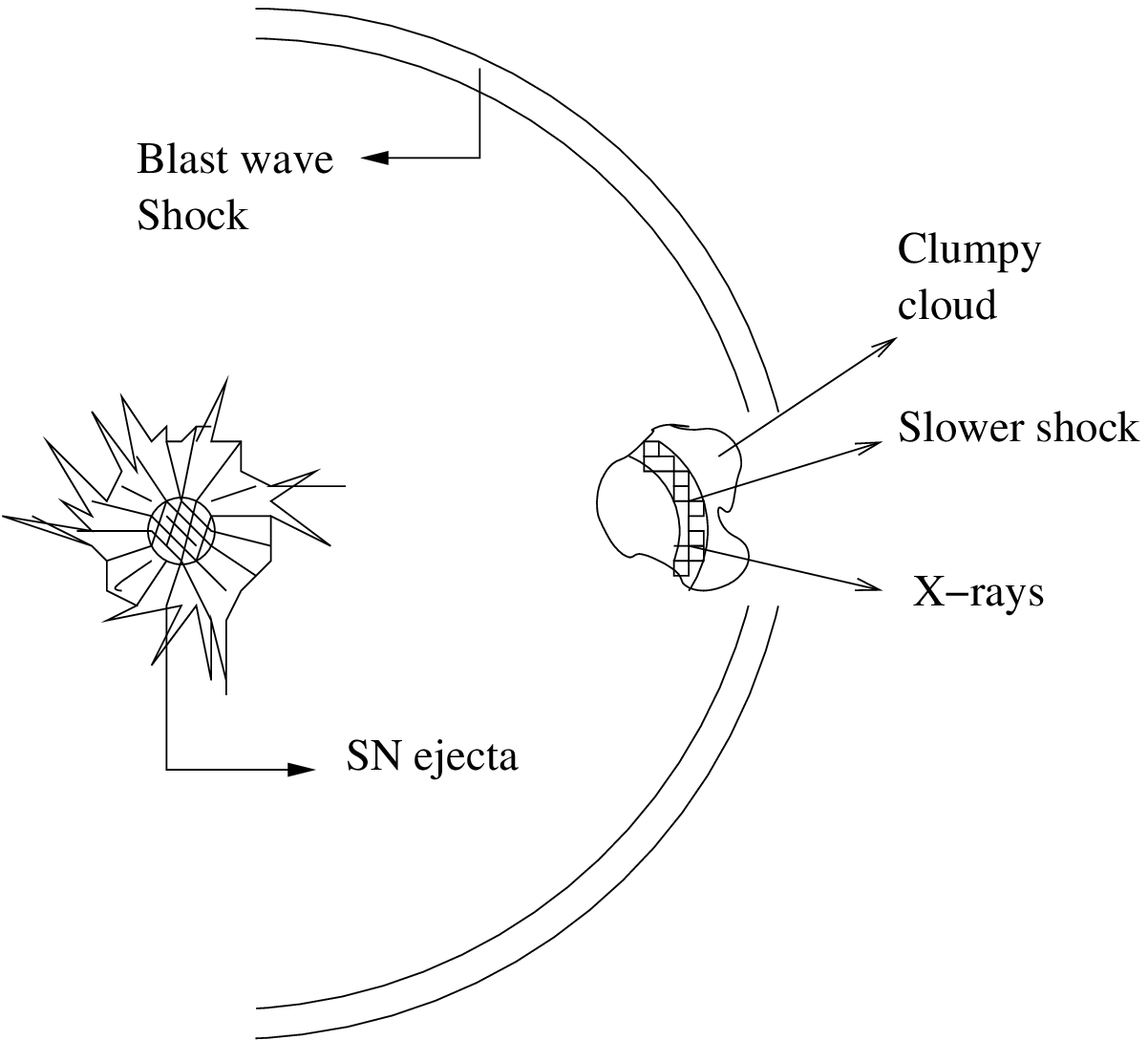}
\caption{Upper panel: supernova ejecta interaction with progenitor wind. 
Reverse shock scenario of X-ray emission\citep{fra02}. 
Ahead of the reverse-shock is the cool shell, which absorbs X-rays.
Lower panel: Interaction of 
supernova ejecta with a cloud in the CS wind \citep{chu93}. When the blast
wave hits the clumpy cloud, it slows down and the X-rays arise from the
slower shock. Figures not to scale. 
\label{models}}
\end{figure}

\begin{figure}
\centering
\includegraphics[angle=270,scale=0.67]{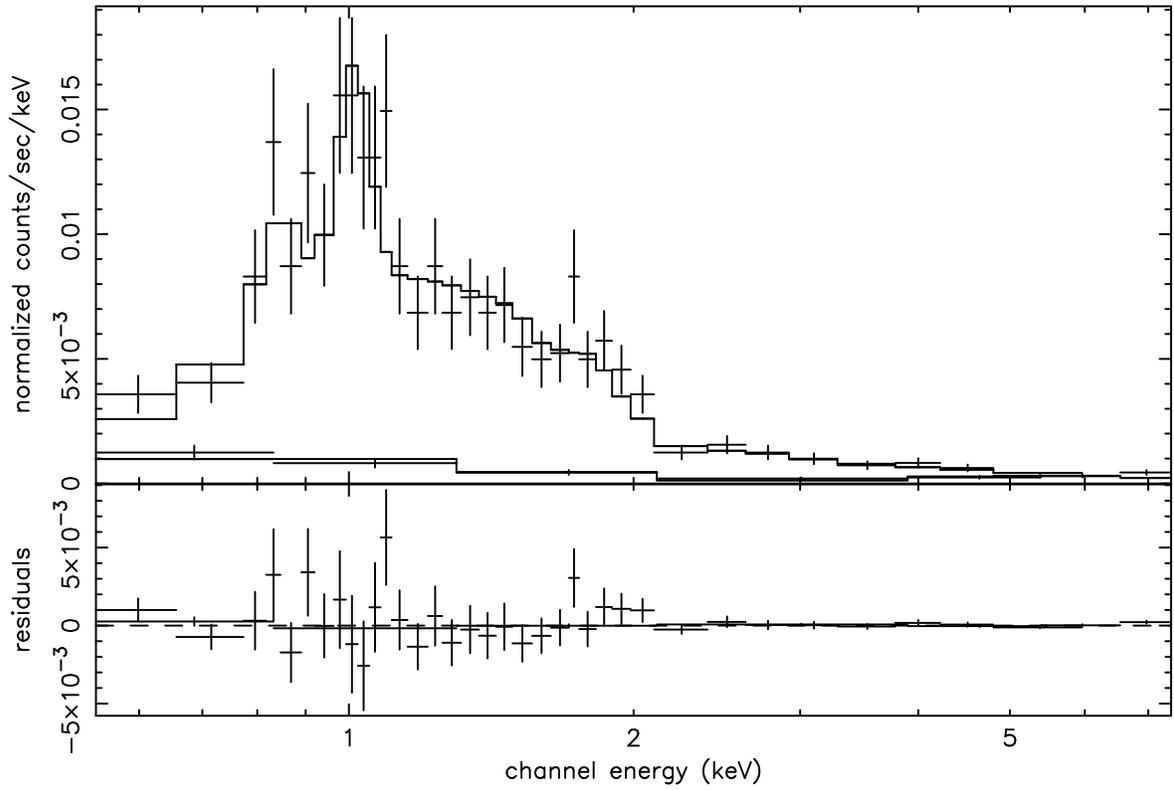}
\caption{
Best fit bremsstrahlung spectra ($kT=2.35$ keV, $N_H=1.51 \times
10^{21}\,{\rm cm^{-2}}$ ) with 2 Gaussian component models to Chandra
data,
with line centroids at $kT_1=1.02\,{\rm keV}$ and $kT_2=0.85\,{\rm keV}$.
The broad horizontal bars in the upper panel represents the 
background levels.
\label{bremsstrahlung} }
\end{figure}

\begin{figure}
\centering   
\includegraphics[angle=270,scale=0.67]{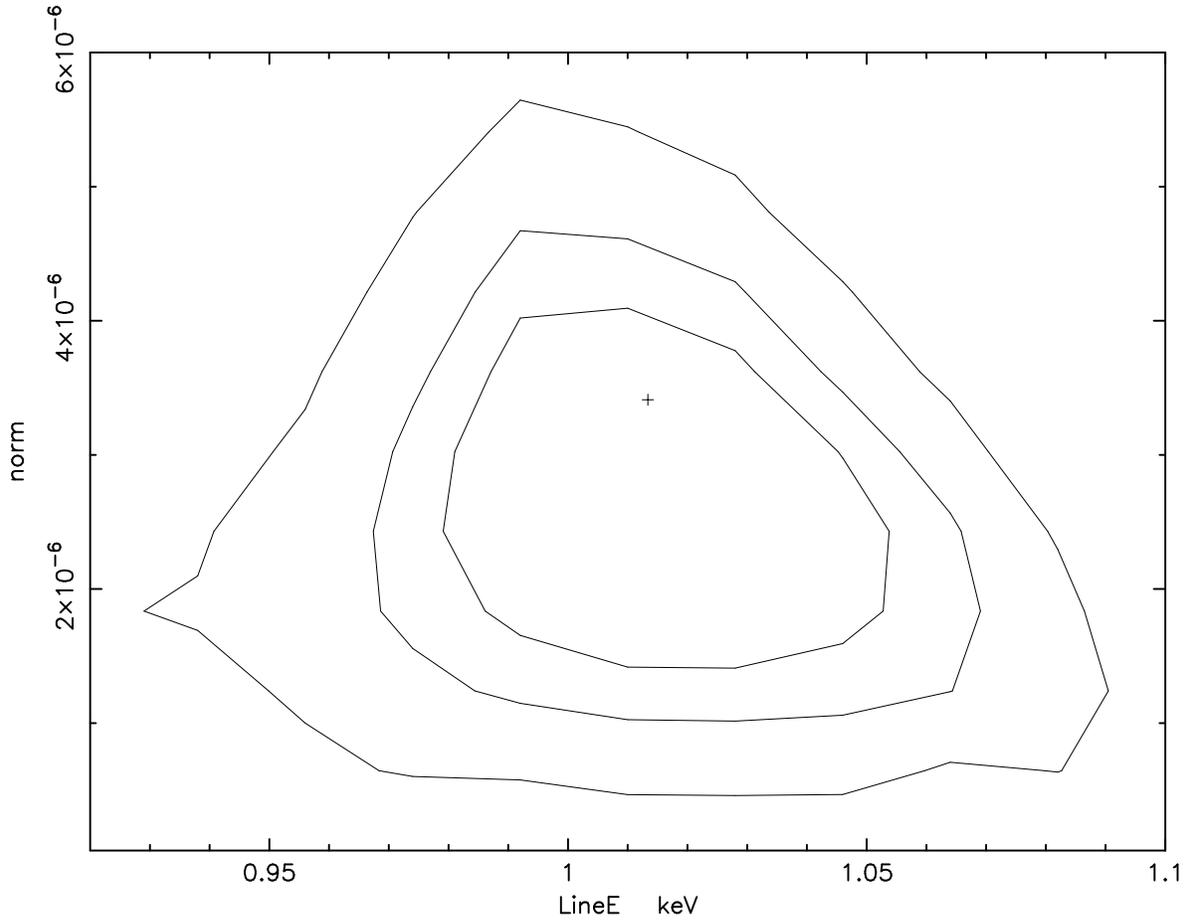}
\caption{ Confidence contours (line energy versus normalization) 
for the line at 1.02 keV (Ne~ X).  
The first contour is at 67\% level, 
the second at 90\% level and the third one is
at 99.9\% confidence level.
\label{fig:bremss+gaus1}}
\end{figure}

\begin{figure}
\centering   
\includegraphics[angle=270,scale=0.67]{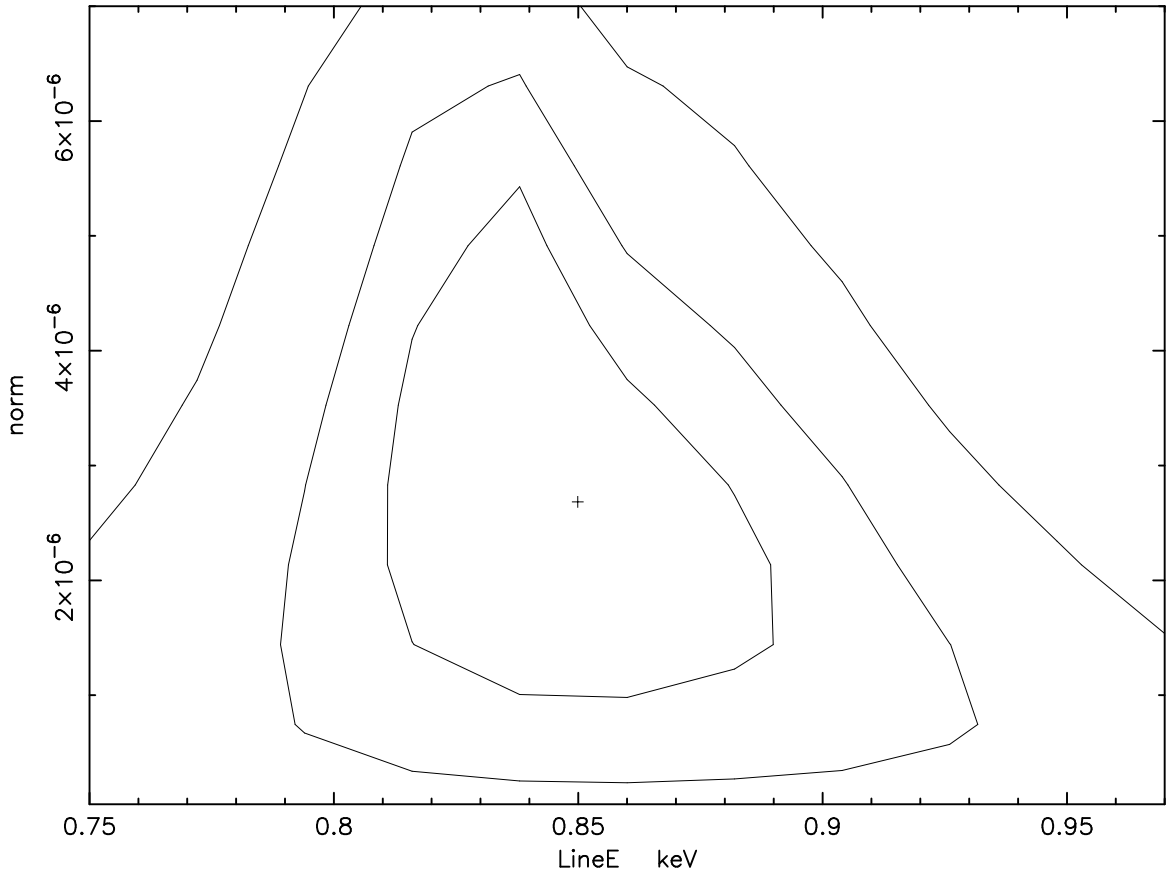}
\caption{ Confidence contours (line energy versus normalization) 
for the line at 0.85 keV (Ne~ IX). 
The first contour is at 67\% level, 
the second at 90\% level and the third one is
at 99.9\% confidence level. The line is detected at 90\%
confidence level.
\label{fig:bremss+gaus+gaus1}}
\end{figure}

\begin{figure}
\centering
\includegraphics[angle=270,scale=0.67]{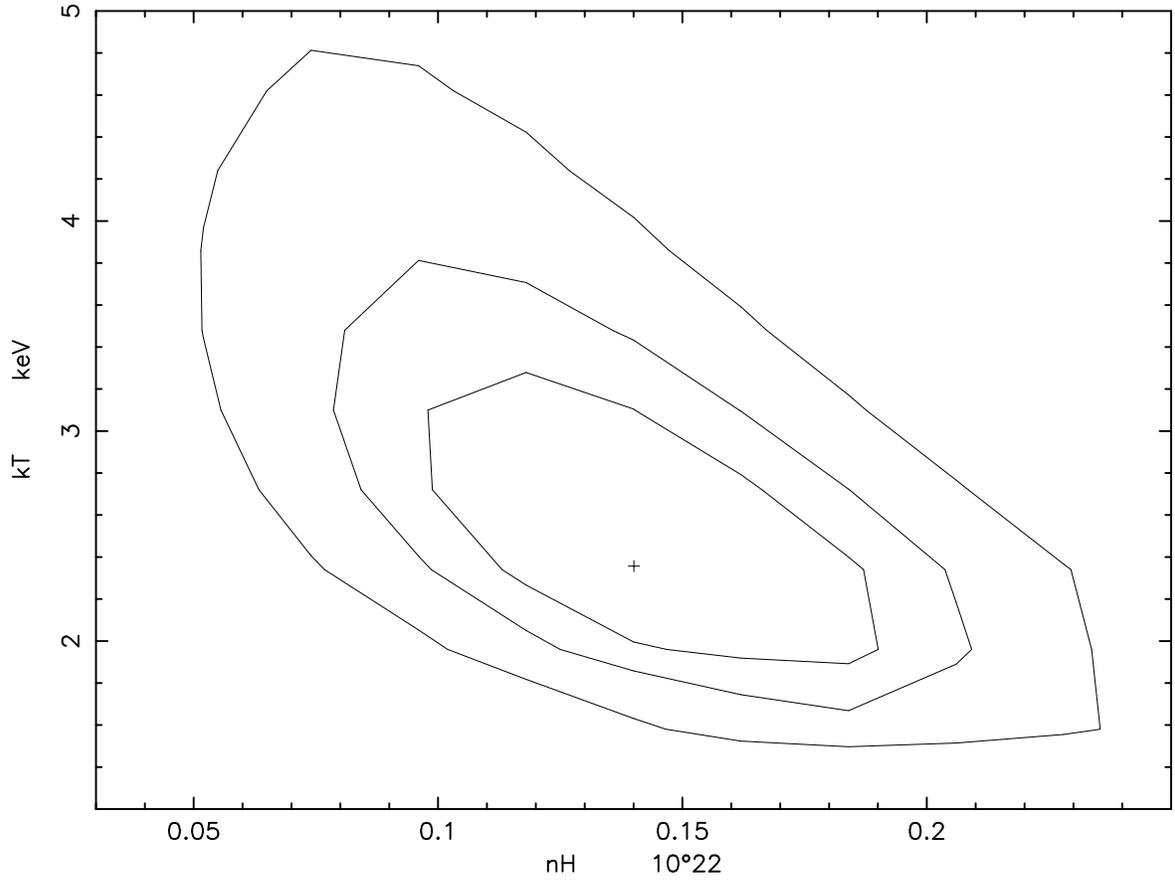}
\caption{Confidence contours of the galactic absorption ($N_H$) and 
the bremsstrahlung temperature ($kT$). 
The first contour is at 67\% level, the second at 90\% 
level and the third one is
at 99.9\% confidence level.
\label{nh_kt} }
\end{figure}

\begin{figure}
\centering
\includegraphics[angle=270,scale=0.67]{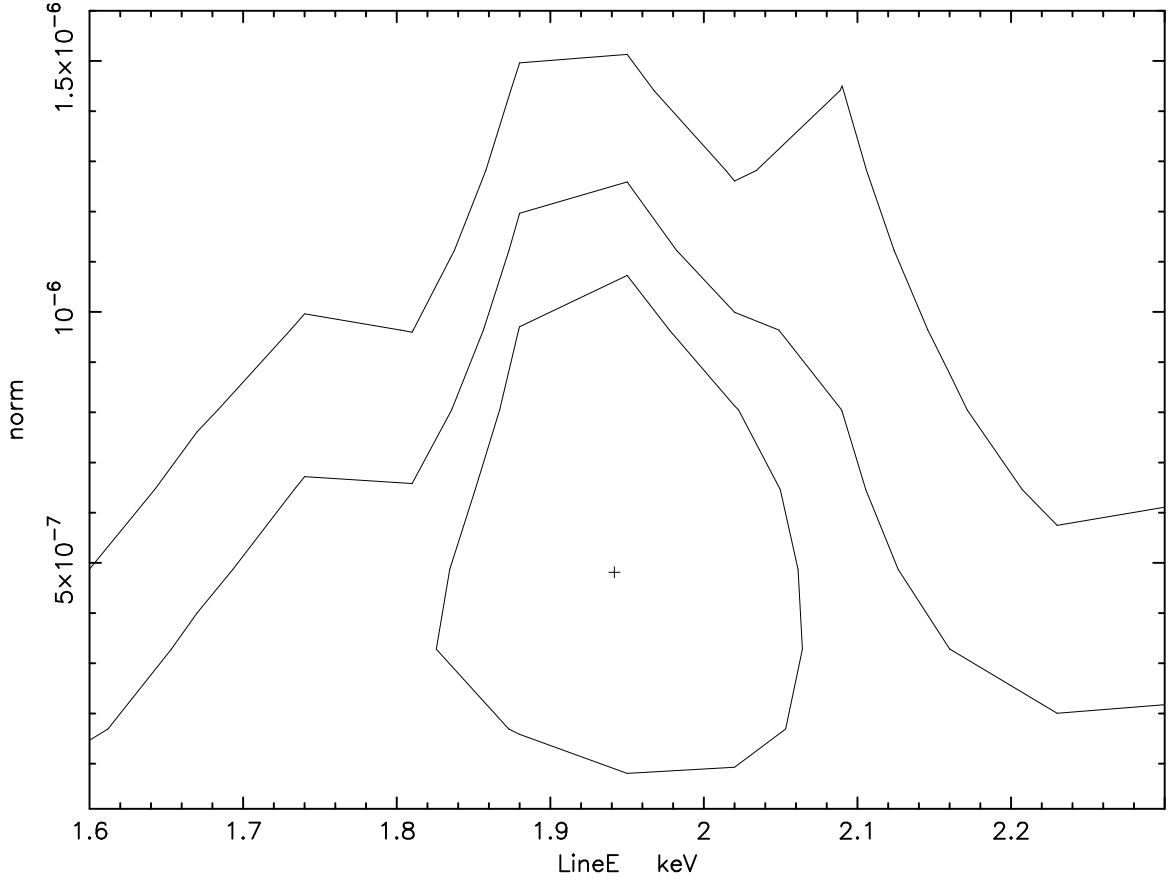}
\caption{ Confidence contours (line energy versus normalization)
for the Si.
The first contour is at 67\% level,
the second at 90\% level and the third one is
at 99.9\% confidence level. Note that there is no lower bound at 90\% 
confidence level, which rules out the possibility of the line.
\label{Si}}
\end{figure}

\begin{figure}
\centering
\includegraphics[angle=0,scale=0.90]{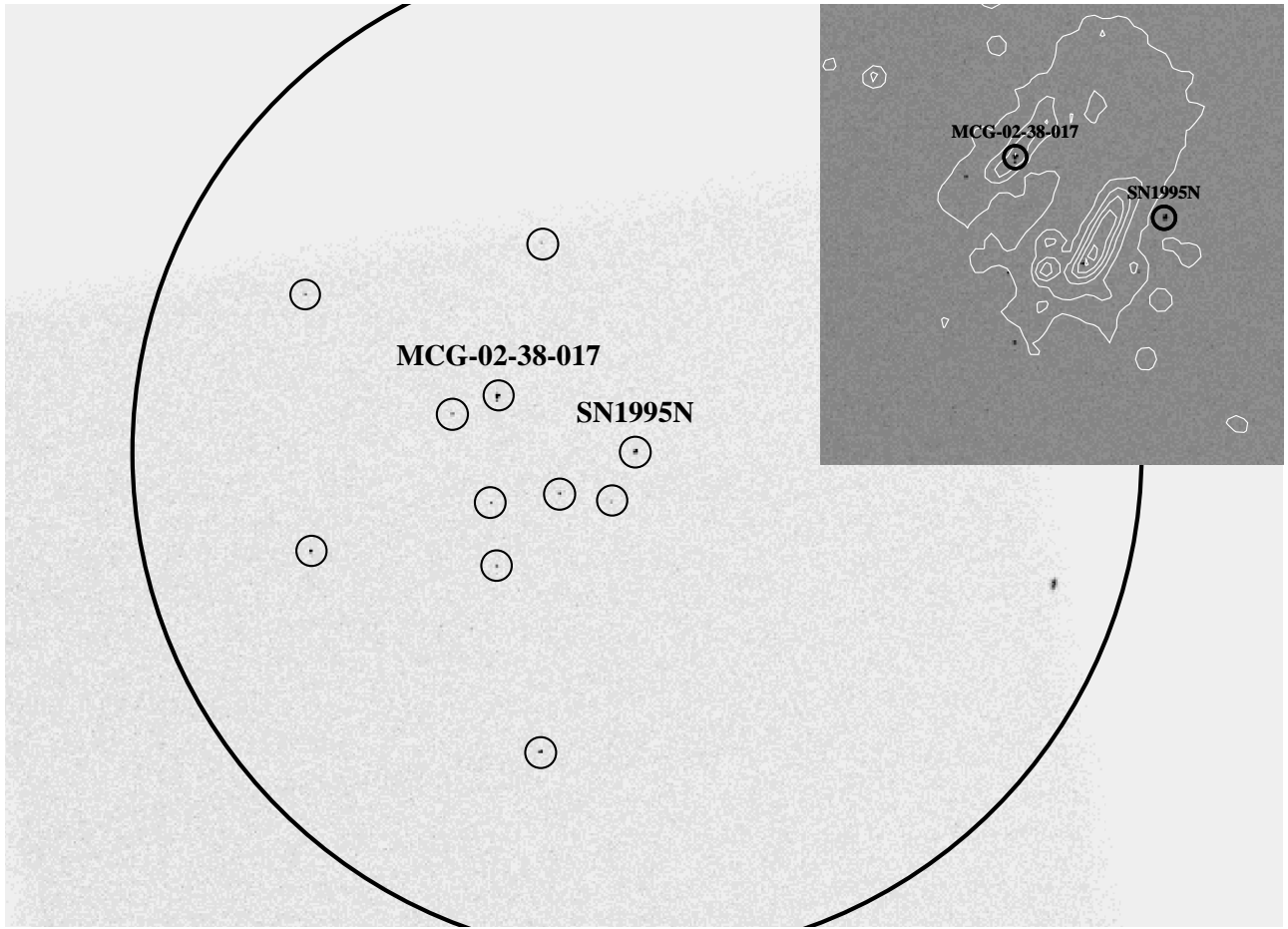}
\caption{
Field of View of SN 1995N for only the ACIS-S3 chip.
In this image, the sources within  a circle of 4$\arcmin$ radius 
are displayed. This was the
size chosen by \citet{fox00} to extract the supernova counts in the ASCA data. 
They had subtracted the
counts of only the right most source ``A" (spot not-circled 
around "3 o'clock") but all the
other sources (shown in circles), including the parent galactic nucleus
must have contributed to the ASCA flux.
The inset shows a grey scale Chandra ACIS-S3 image overlaid
with the optical contours obtained from the Digital Sky Survey (DSS).
\label{fov}}
\end{figure}

\begin{figure}
\centering
\includegraphics[angle=270,scale=0.45]{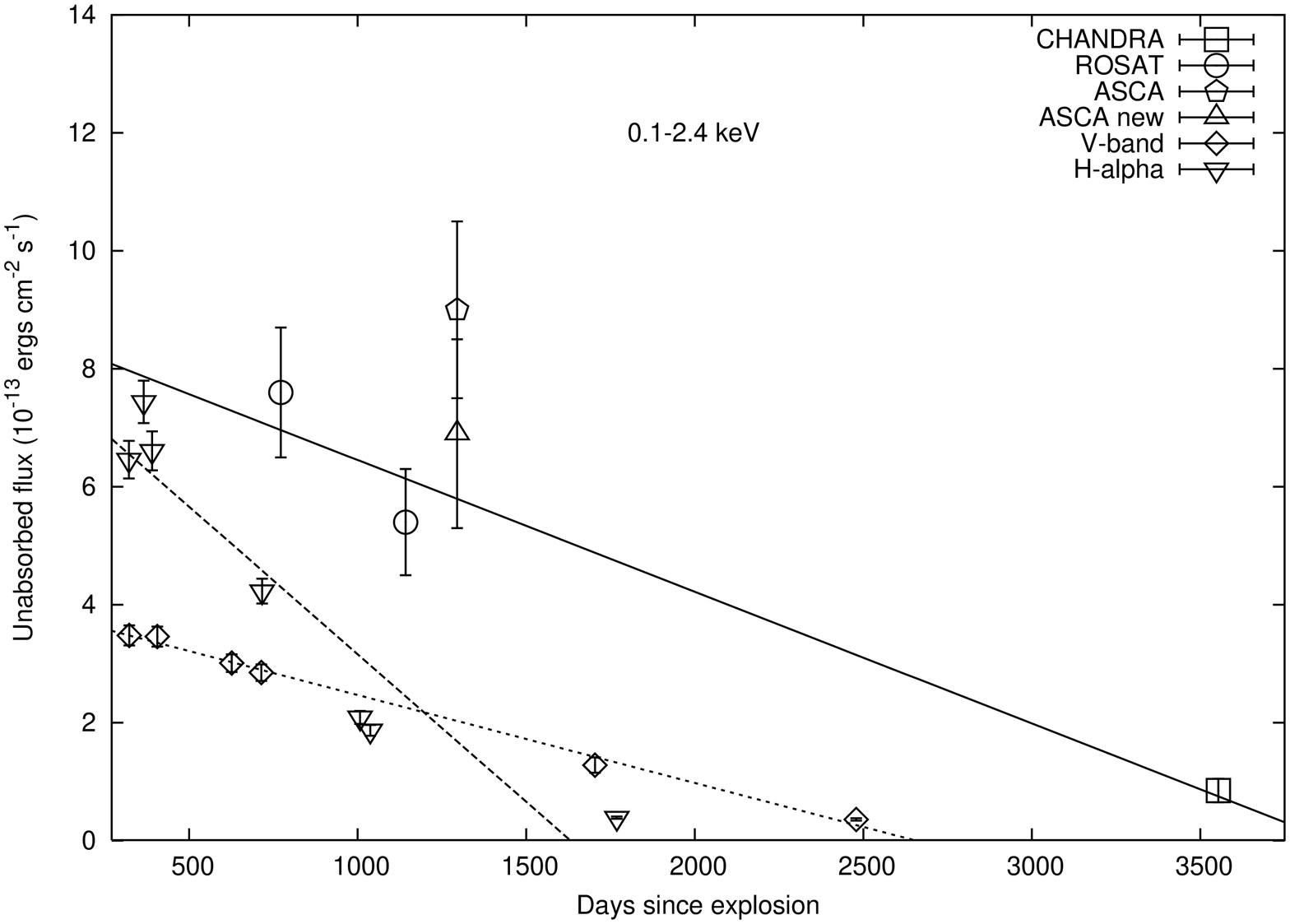}
\includegraphics[angle=270,scale=0.45]{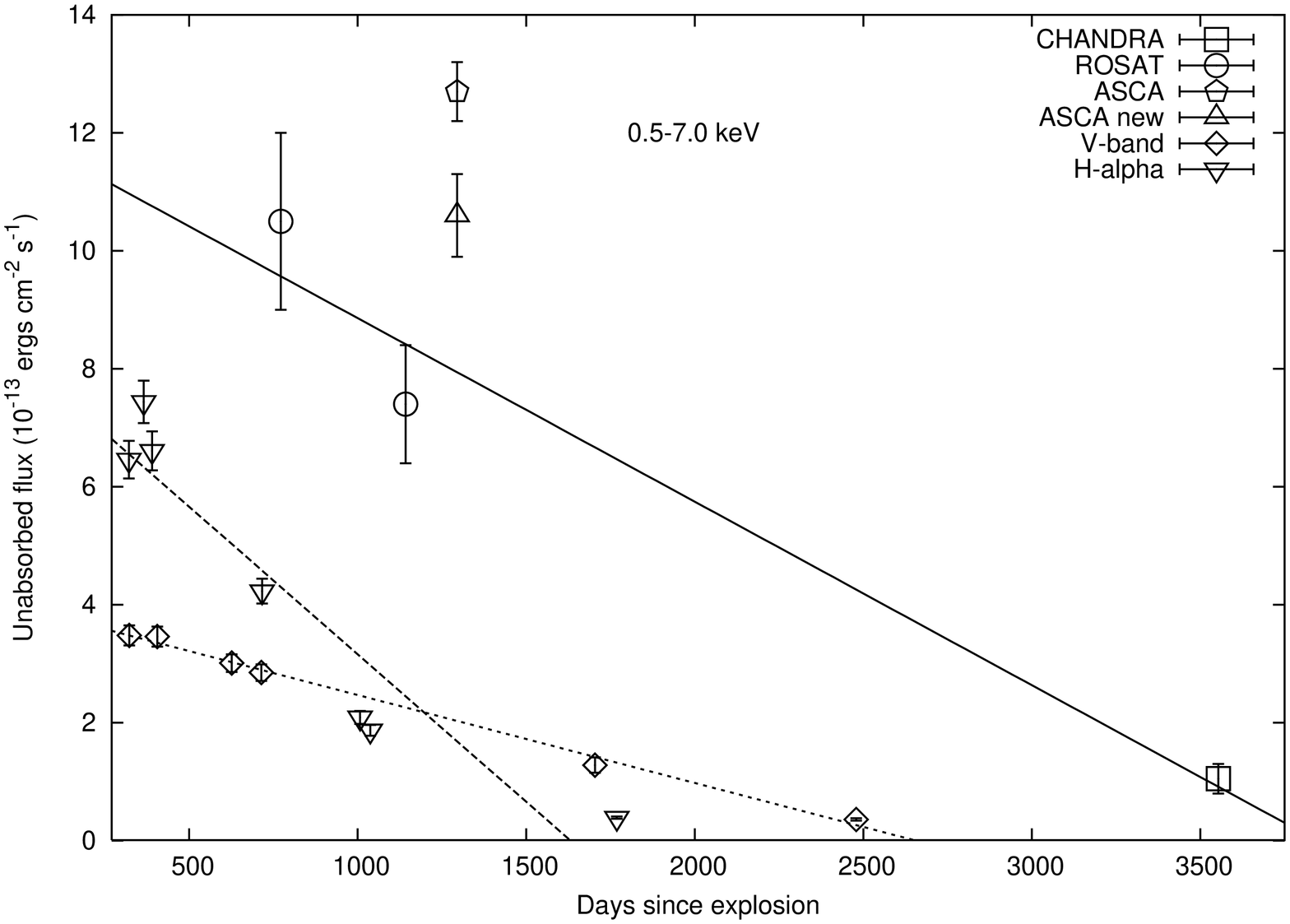}
\caption{
The upper panel shows the unabsorbed X-ray 
light curve for 0.1-2.4 keV. The lower panel shows the corresponding light
curve for the 0.5-7.0 keV band. 
Solid lines are fits to the ROSAT and the Chandra data. The data point with
upper-triangle symbol shows the 
revised ASCA points after taking out the contribution from the other
point sources in the 4$\arcmin$ Field of View. 
Dashed line shows the de-reddened optical V-band light curve \citep{li02}. 
Dot-dashed line shows
the evolution of reddening-corrected H-$\alpha$ flux \citep{fra02}. 
Note that SN 1995N was
discovered $\sim 10 $ months after the explosion and 305 days have been
added to the observation dates to arrive at abscissae in the above Figures.
\label{new_lc}}
\end{figure}

\begin{figure}
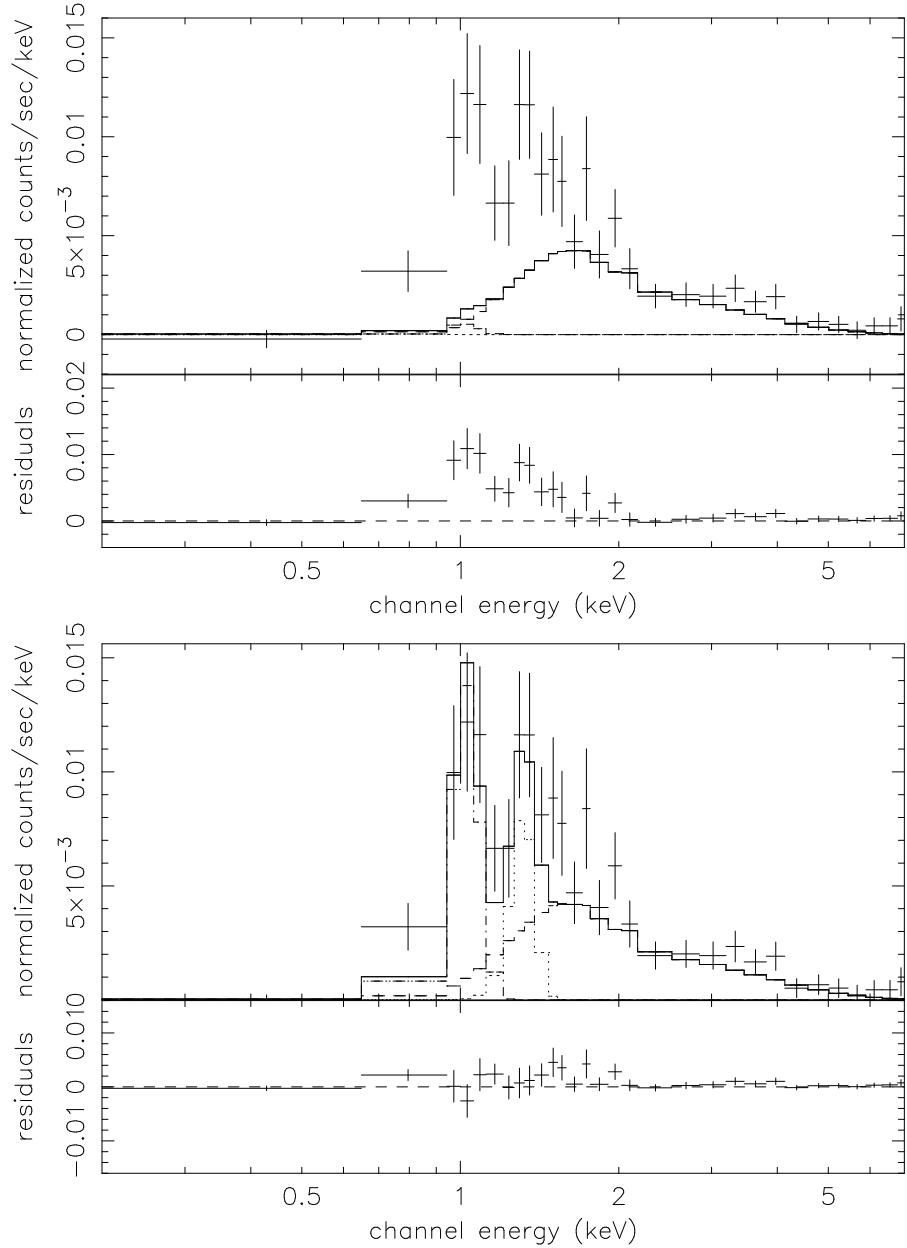

\centering
\includegraphics[angle=270,scale=0.5]{f9a.eps}
\includegraphics[angle=270,scale=0.5]{f9b.eps}
\caption{
Upper panel: The ASCA spectrum with best fit Chandra model. 
It shows a significant difference between the
two because of the contamination
and possible evolution. 
Lower panel:
The best-fit bremsstrahlung model to the ASCA data, yielding 
: $N_H=1.31 \times 10^{21}\,{\rm cm^{-2}}$, 
and $kT=2.53$ keV, with lines at
1.02 and 1.32 keV.  
\label{fig:asca_bremss} }
\end{figure}

 \citet{fra96} have shown that when the ejecta gradient is moderately
flat ( $n < 8$),
both shocks (circumstellar and reverse) are adiabatic and most flux below
10 keV comes from the reverse shock.
Luminosity from the reverse shock can be expressed as \citep{che03}:
$$L_{rev}=2 \pi R_s^2 \rho_{ej} V_{rev}^3$$
Since the reverse shock velocity
$V_{rev} \propto t^{-1/(n-2)}$, this means that the 
total luminosity of the reverse shock decreases linearly with time: 
$$L_{rev} \propto t^{3-2s}\propto 1/t$$
Therefore, the observed linear decline suggests that, the 
lower temperature
ejecta gas struck by the reverse-shock
can account for the soft X-ray emission
seen from young supernovae (see Fig. \ref{models}, upper panel) where the
velocity scale (and line-width) of $\sim 10,000 \rm \;km \; s^{-1}$
is set by the expanding stellar ejecta \citep{che03}.   
The alternate model of \citet{chu93}, in which the soft X-rays can emerge 
from the radiative cooling of  shocked, dense clumps 
embedded in the circumstellar wind
overtaken by the blast-wave shock and crushed by the
shocked wind (Fig. \ref{models}, lower panel), 
would require a time dependent turn-on of the shocked clouds. 
 In view of the steady decline seen in SN 1995N, this model is
 less likely 
to be valid unless the sampling of the light curve has been 
infrequent enough to miss out bumpy features due to the large number
of small clouds.

\subsection{Neon ejecta and implications for stellar models}

It is well known that major source of galactic supply of $^{12}{\rm C}$
and $^{16}{\rm O}$ stars are the Red Giant stars burning He to 
produce these major
ashes. Fortuitous circumstances of the energy level parameters
of these $\alpha-$particle nuclei are important for the observed
abundance of Oxygen and Carbon (see e.g. \citet{rol88}), where 
$^{16}{\rm O}$ producing Neon via 
$^{12}{\rm C(\alpha,\gamma)^{16}O}$, is not
wholly burnt away by $^{16}{\rm O(\alpha,\gamma)^{20}Ne}$.
This is however not a universal constraint
as higher core temperatures expected in Supergiant stars
can broaden the Gamow Peak window so that more
channels in the final $^{20}{\rm Ne}$ nucleus open-up so that the overall
astrophysical reaction rates are substantially 
increased over those prevalent in Red Giants. As a result, the final 
nucleosynthetic output from, say, a 25 M$_{\odot}$ supernova
may even yield   dominant production factors of Neon with
respect to solar Neon 
isotopes compared to those of Oxygen.
In Table \ref{woosley}, we provide a summary of the dominant elements
in certain interior mass ranges obtained from the final composition by 
mass fraction of two presupernova stars of main-sequence mass 
$15 M_{\odot}$ and $25 M_{\odot}$ as provided in Fig. 9 of  \citet{woo02}.
The inner mass ranges reported in this Table have O, Ne and Mg
cores and these elements are products of core C-burning or of 
partially consumed C-shell burning. The outermost layers reported
in this Table however have substantial $^{20}{\rm Ne}$ but are practically
devoid of both O and Mg and are products of incomplete
He shell-burning. It is  plausible that the Ne lines seen
in the Chandra data originate in these shells. 
Other elements in these shells, like C, N or He do not have X-ray
lines in an energy range where ACIS-S has substantial sensitivity. Since
the He- and H-rich layers form a part of a high velocity gas seen in the
broad spectral components of the optical and UV spectra of SN 1995N,
if the Ne that we observe is mixed in with the Helium layers, then most
likely, the Ne X-ray lines are also arising among the same broad-line
component. Alternatively, if the Ne arises together with O and Mg
in an O-Ne-Mg core as a result of core C-burning, one would
normally also expect lines of O and Mg to be present in the Chandra
X-ray spectrum. However, we note that due to the poor counting statistics,
we are already near the threshold of detection of the strongest line,
that of Ne, and the absence of O or Mg lines could be due to the
 lower sensitivity of the ACIS at these energies.
The possibility that Neon emission arises from the O-Ne-Mg core
is however unlikely, as Table \ref{woosley} shows that successively
higher amounts of Neon are overproduced in the inner zones, compared to
the outer layers. 

Among the isotopes of Neon, $^{20}{\rm Ne}$ and $^{21}{\rm Ne}$
are primarily products of Carbon burning as also are $^{24,25,26}{\rm Mg}$
(see \citet{woo02} Table 3), whereas $^{22}{\rm Ne}$ (together
with $^{16}{\rm O}$ and  $^{18}{\rm O}$) are products of He-burning
in nucleosynthesis resulting from  massive
stars with $11-40 M_{\odot}$ and various metallicities. In fact, the 
dominant Ne isotope for a solar metallicity 
star of $25 M_{\odot}$ at the end of the He-burning is $^{22}{\rm Ne}$. 
The ACIS-spectrum is unable to distinguish between different
isotopes of the same element. Therefore, a fraction of the Neon seen from
the SN 1995N spectrum could be due to $^{22}{\rm Ne}$, especially if the 
supernova arose from a more massive progenitor. $^{22}{\rm Ne}$ is made from 
$^{18}{\rm O}$ at high temperatures in reactions 
$^{18}{\rm O(\alpha,\gamma)^{22}Ne}$ during the He-burning. The 
more recent measurements of the $^{18}{\rm O(\alpha,\gamma)^{22}Ne}$
reaction \citep{gie93} indicate that this reaction rate may be much higher
than that of \citet{cau88} in which case most of the $^{18}{\rm O}$
will end up in $^{22}{\rm Ne}$. 
The neutron-rich seed 
nucleus $^{18}{\rm O}$ is in turn made in massive stars in 
the sequence in the reaction 
$^{14}{\rm N(\alpha,\gamma)^{18}F(e^+,\nu)^{18}O}$.
Thus $^{22}{\rm Ne}$ comes effectively from two $\alpha-$captures on the
$^{14}{\rm N}$ left over from the CNO cycle (during the Hydrogen
burning phase) and the amount of $^{22}{\rm Ne}$ scales linearly
with the initial metallicity of the star \citep{woo02}. 
$^{22}{\rm Ne}$ itself would be destroyed due to
$^{22}{\rm Ne(\alpha,n)^{25}Mg}$ reactions in the high temperature 
s-process occurring late in the Helium burning stage.
The fact that the Chandra spectra reveals Ne lines indicates that either the 
progenitor of SN 1995N was not sufficiently
massive to destroy $^{22}{\rm Ne}$ in this manner  or the 
initial metallicity of the  star was not negligible or that 
the $^{22}{\rm Ne(
\alpha,n)^{25}Mg}$ rate may have been overestimated. 

\section{Conclusions}

Below we summarize the main conclusions of this paper:

\begin{itemize}

\item  The Chandra spectrum of SN 1995N is different from the spectrum
of the same
region observed by ASCA in 1998 that we have reanalyzed here, especially in 
the soft energy bands. We detect a Ne X line in both observations, and while
we detect a Ne IX line in the Chandra observation this was absent in
the ASCA observation. At the same time we detect a 1.3 keV line in the
ASCA observation that is absent in the Chandra spectrum of SN 1995N.
No Fe line was detected in either spectrum. Fe, if present is in a cold state,
 without having undergone significant mixing with outer layers.

\item After taking out the contribution from the contaminating sources
in ASCA PSF,  the light curve appears to be consistent with
a linear decline.  This indicates that the X-ray emission is
due to the reverse shock going through a shallow ejecta profile.

\item The observed absorbed column depth seem to indicate an extra 
component over and above due to the galactic column absorption.
It is likely to be due to a thin cool shell between reverse-shock and 
the contact discontinuity,  as discussed in the previous sections.

\item  About 0.01 $M_\odot$ of Ne in SN 1995N 
is estimated from the Chandra line  detection,
which most likely, arises in the partially burnt He core, at velocities, 
greater than 5000 km s$^{-1}$.

\end{itemize}

\acknowledgments

We thank the Chandra X-ray Observatory team for carrying out
the  project and support from the Chandra
Guest Observer program through a NASA grant. 
PC thanks the hospitality of Center for Astrophysics where
most of the data analysis was carried out.
She is an awardee of the Sarojini
Damodaran International Fellowship and 
the Kanwal Rekhi Career Development award.
FKS is supported by the Alexander von Humboldt Foundation.
The research of EMS is supported by contract NAS8-39073 to SAO to
operate the Chandra Observatory. 
At Tata Institute this research is a part of Project No. 10P-201 of
the Tenth Five Year Plan.  One of the authors (PC) thanks Tanja Nymark 
for discussing her work in advance of publication.
We thank the anonymous referee for his/her comments and suggestions
to improve the presentation and clarity of the manuscript.

\clearpage
 
\begin{deluxetable}{lllcccccc}
\tabletypesize{\scriptsize}
\tablewidth{455pt}
\tablecaption{Summary of all X-ray observations of SN 1995N
\label{all_obs}}
\tablehead{
\colhead{Date} & \colhead{Mission/Inst.} & \colhead{Exp.} & \colhead{Counts} &
\colhead{kT} & \colhead{$N_H$\tablenotemark{b}} &
\multicolumn{2}{c}{Unabsorbed flux\tablenotemark{c}}  &
\colhead{$L_X$\tablenotemark{d}} \\
\cline{7-8}
\colhead{} &
\colhead{} &
\colhead{ks}  & \colhead{} & \colhead{keV} & \colhead{} &
\colhead{0.1-2.4 keV} &
\colhead{0.5-7.0 keV} &
\colhead{$10^{40}\,{\rm erg\,s^{-1}}$} 
 }
\startdata
1996 Jul-Aug & $ROSAT$ HRI & 18.30 & 172 & $9.1^{+2.7}_{-1.8}$\tablenotemark{a} 
& $1.1 \pm 0.4$\tablenotemark{a} & $6.5-8.7$ & $9.0-12.0$ & 12 \\
1997 Aug 17  & $ROSAT$ HRI & 18.80 & 126 & $9.1^{+2.7}_{-1.8}$\tablenotemark{a}
& $1.1 \pm 0.4$\tablenotemark{a} & $4.5-6.3$ & $6.4-8.4$ & 8 \\
1998 Jan 20  & $ASCA$ SIS & 91.13 & 1960 & $9.1^{+2.7}_{-1.8}$ &
$1.1 \pm 0.4$  & $7.5-10.5$ & $12.2-13.2$ & 14 \\
             & $ASCA$ GIS & 95.94 & 1300 & $9.1^{+2.7}_{-1.8}$
& $1.1 \pm 0.4$ & $8.6-11.5$ & $14.0-15.2$ & 15 \\
2004 Mar 27  & $CHANDRA$  & 55.74 & 758 & $2.35^{+1.45}_{-0.75}$ &
$1.51^{+0.62}_{-0.64}$ & $0.65-1.05$ & $0.80-1.30$ & 2.3\\
\enddata
\tablenotetext{a} {Obtained from spectral fits of {\it ASCA} data since 
{\it ROSAT} HRI does not have spectral response.}  
\tablenotetext{b} {N$_H$ is in units of $10^{21}\, {\rm cm^{-2}}$}.
\tablenotetext{c} { Flux in $10^{-13}\,{\rm erg\,cm^{-2}s^{-1}}$. 
Bremsstrahlung model with the indicated ($kT,\, N_H$)
is used to calculate the flux.}
\tablenotetext{d} { Unabsorbed Luminosities  
are in the energy range $0.1-10.0$ keV band; d=24 Mpc.}
\tablenotetext{e} { Reference for ROSAT and ASCA measurements is \citet{fox00}}
\end{deluxetable}

\clearpage

\begin{deluxetable}{ccccc}
\tabletypesize{\footnotesize}
\tablewidth{435pt}
\tablecaption{Best fit parameters for the broken powerlaw  model
for the background.
 \label{bkd1}}
\tablehead{
\colhead{Photon index 1} & \colhead{Photon index 2} & \colhead{Break Energy (keV)} & \colhead{Normalization} & \colhead{${\chi}^2/d.o.f.$} 
}
\startdata
$2.465 \pm 0.428$ & $-2.50 \pm 1.00$ & $2.667\pm0.231$ & $(2.58\pm0.69) \times 10^{-6}$ & $2.37/3 d.o.f$\\
\enddata
\end{deluxetable}

\clearpage

\begin{deluxetable}{cccccccc}
\tabletypesize{\scriptsize}
\tablewidth{445pt}
\tablecaption{Spectral Model Fits to ACIS Spectrum\tablenotemark{a}
\label{bestfit_models}}
\tablehead{
 \colhead{Model} & \colhead{${\chi}^2/{\nu}$} & \colhead{DoF} & 
\colhead{N$_{\rm H}$}  & \colhead{Param-1} &
 \colhead{Param-2} & \colhead{Eq. Width} & \colhead{Norm ($10^{-5}$)} }
\startdata
Brems & 1.61 & 35 & $1.01\pm0.04$ & kT=$2.20\pm0.41$ &
$\cdots$ & $\cdots$ & $3.64\pm1.09$ \cr
\hline
Brems + & 1.09 & 33 & $1.51^{+0.62}_{-0.64}$ & 
kT=$2.35^{+1.51}_{-0.75}$ & $\cdots$ & $\cdots$ & $3.50\pm1.70$
\cr
~~+ Gauss\tablenotemark{b} & $\cdots$ & $\cdots$ & $\cdots$ & E$_{\rm line}=
1.02^{+0.04}_{-0.04}$ & 
N\tablenotemark{c}=$0.34\pm0.19$ & $129^{+83}_{-20} $ eV& $\cdots$ \cr
~~+ Gauss\tablenotemark{b} & $\cdots$ & $\cdots$ & $\cdots$ &E$_{\rm line}=
0.85^{+0.08}_{-0.04}$ &
N\tablenotemark{c}=$ 0.27\pm0.41$ & $<200$ eV & $\cdots$ \cr
\hline
VMekal & 1.51 & 33 & $0.37\pm0.27$ & 
kT=$3.03^{+0.36}_{-0.34}$ & $\cdots$ & $\cdots$ & $5.21\pm0.97$ \cr
+ Ne & $\cdots$  & $\cdots$  & $\cdots$  & 
abun\tablenotemark{d}=$10.12\pm4.18$ & $\cdots$
& $\cdots$ & $\cdots$ \cr 
+ Si & $\cdots$  & $\cdots$  & $\cdots$  & 
abun\tablenotemark{d}=$4.03\pm1.6$ & $\cdots$
& $\cdots$ & $\cdots$ \cr 
\hline
Brems + & 1.20 & 34 & 0.61\tablenotemark{e} 
& kT=$3.17\pm0.60$ & $\cdots$ & $\cdots$ & $2.62\pm0.71$
\cr
~~+ Gauss\tablenotemark{b} & $\cdots$ & $\cdots$ & $\cdots$ 
& E$_{\rm line}=1.02\pm0.02$ & 
N\tablenotemark{c}=$0.29\pm0.16$ & $124^{+82}_{-70}$ eV &  $\cdots$ \cr
~~+ Gauss\tablenotemark{b} & $\cdots$ & $\cdots$ 
& $\cdots$ &E$_{\rm line}=0.86\pm0.06$ &
N\tablenotemark{c}=$ 0.13\pm0.22$ & $< 135$ eV & $\cdots$ \cr
\enddata
\tablenotetext{a}{All errors listed are 90\%.  
Units are 10$^{21}$ cm$^{-2}$ for
 N$_{\rm H}$, keV for kT; keV for line energy.}
\tablenotetext{b} {Gaussian widths were fixed to zero.}
\tablenotetext{c} {Normalization of the Gaussian component 
in units of $10^{-5}$.} 
\tablenotetext{d} {Abundances are with respect to solar abundances: 
${\rm (X/X_{\odot}})$}
\tablenotetext{e} {Galactic absorption $N_H$ was fixed to $6.1 \times
10^{20}\,{\rm cm^{-2}}$. Obtained from the galactic extinction calculation.} 
\end{deluxetable}

\clearpage

\begin{deluxetable}{cccccc}
\tablewidth{480pt}
\tablecaption{Upper limits on the various line strengths and their abundances,
in VMekal models.
\label{other_lines} }
\tablehead{
\colhead{Element} & \colhead{Line Energy} &
\multicolumn{2}{c}{Normalization} & \multicolumn{2}{c}{Abundance wrt Solar} \cr
\cline{3-4}
\cline{5-6} 
\colhead{} & \colhead{keV} & \colhead{Central Value} & \colhead{Error Range} & 
\colhead{Central Value} & \colhead{Error Range} }
\startdata
O & 0.65 & $7.73\times10^{-10}$ & $0-5.70\times10^{-6}$ & $0.01$ & $0-0.44$\\
Mg & 1.36 & $2.16\times10^{-10}$ & $0-5.67\times10^{-7}$ & $0.5$ & $0-2.8$\\
Si & 1.86 & $2.45\times10^{-7}$ & $ 0-1.42\times10^{-6}$ & $4.0$ & $2.4-5.6$\\
S & 2.4 & $3.15\times10^{-8}$ & $ 0-8.47\times10^{-7}$ & $2.2$ & $0-4.8$\\
Fe & 6.7 & $8.16\times10^{-9}$ & $ 0-1.85\times10^{-4}$ & $0.05$ & $0-0.30$\\
\enddata
\end{deluxetable}

\clearpage

\begin{deluxetable}{lcccccc}
\tabletypesize{\scriptsize}
\tablewidth{473pt}
\tablecaption{Absorbed fluxes 
predicted in best-fit VMekal and Bremsstrahlung  plus lines models.
\label{fluxes} }
\tablehead{
\colhead{Model} & 
\multicolumn{6}{c}{Fluxes}\cr 
\cline{2-7}
\colhead{} & 
\multicolumn{2}{c}{0.3-7.5 keV} & 
\multicolumn{2}{c}{0.1-2.4 keV} & 
\multicolumn{2}{c}{0.5-7.0 keV} \cr
\cline{2-3}
\cline{4-5}
\cline{6-7}
\colhead{} & 
\colhead{$10^{-5}\,{\rm  cm^{-2}/s}$} &
\colhead{$10^{-13}\,{\rm erg\, cm^{-2}/s}$} & 
\colhead{$10^{-5}\,{\rm  cm^{-2}/s}$} & 
\colhead{$10^{-13}\,{\rm erg\, cm^{-2}/s}$} & 
\colhead{$10^{-5}\,{\rm cm^{-2}/s}$} & 
\colhead{$10^{-13}\,{\rm erg\, cm^{-2}/s}$} 
}
\startdata
Bremss & $2.97\pm0.71$ & $0.76\pm0.18$ & $2.46\pm0.58$ & $0.47\pm 0.11$ & $2.47\pm0.59$ & $0.75\pm0.18$\\ 
VMekal &  $3.62\pm0.87$ & $0.90\pm0.20$ & $2.99\pm0.72$ & $0.50\pm 0.12$ 
& $3.14\pm0.75$ & $0.85\pm0.19$\\ 
\enddata
\end{deluxetable}

\clearpage

\begin{deluxetable}{cclllcl}
\tabletypesize{\scriptsize}
\tablewidth{460pt}
\tablecaption{Elements co-synthesized with Ne and their mass fractions 
 in the interiors of massive
stars prior to supernova explosion (after \citet{woo02}, Fig. 9).
\label{woosley}}
\tablehead{
\colhead{M$_{STAR} $}   & \colhead{Mass}  & 
\colhead{Composition} & \colhead{X$_{Ne}$} & 
\colhead{X$_{O}$} & \colhead{Ne mass\tablenotemark{a}} & \colhead{Comments}\\
\colhead{on ZAMS} & \colhead{co-ordinates} &
\colhead{} & \colhead{}&
\colhead{} & \colhead{in the layer} & \colhead{}\\ 
\colhead{(M$_{\odot}$)} & \colhead{(M$_{\odot}$)} & \colhead{}
 &\colhead{} &\colhead{} & \colhead{(M$_{\odot}$)} & \colhead{} }
\startdata
15 M$_{\odot}$ & $(1.8-2.6)$M$_{\odot}$ & $^{16}{\rm O}, 
^{20}{\rm Ne},^{24}{\rm Mg}$ & 0.26 & 0.7 & $2.1 \times 10^{-1}$ & O+Ne+Mg core: Product of\\
 & & & & & &C-burning in C+O core\cr 
 & $(2.6-3.05)$M$_{\odot}$ & $^{16}{\rm O}, 
^{12}{\rm C},^{20}{\rm Ne}, ^{24}{\rm Mg}$ & $\le 0.05$ & 0.75
&$2.3\times10^{-2}$& C-burning around C+O core\cr 
& $(3.05-3.8)$M$_{\odot}$ & $^{4}{\rm He}, 
^{12}{\rm C},^{20}{\rm Ne}, ^{14}{\rm N}$ & 0.0133 & 0.002  &
$1\times10^{-2}$&
Partially burnt Helium in  \\
 & & & & & & He-shell burning\\
 & $(3.8-4.2)$M$_{\odot}$ & $^{4}{\rm He},^{14}{\rm N},
^{20}{\rm Ne} $&0.0017
& 0.008 & $6.8\times10^{-4}$ & Unburnt He-core\cr
& & & & & & \\
\hline
& & & & & & \\
25 M$_{\odot}$ & $(1.9-5.7)$M$_{\odot}$ & $^{16}{\rm O},
^{20}{\rm Ne},^{24}{\rm Mg}$ & 0.2 & 0.7 & $7.6 \times 10^{-1}$ 
& O+Ne+Mg core: Product of\\
 & & & & & & C-burning in part of C+O core\cr
 & $(5.7-7.1)$M$_{\odot}$ & $^{16}{\rm O}, ^{12}{\rm C}, ^{20}{\rm Ne}$ & 
$\le 0.08$ & 0.57 & $1.1\times10^{-1}$ & Part of C+O core: Product of\\
& & & & & & complete He- and C-burning at\\
& & & & & & high temperature\cr
 & $(7.1-8.1)$M$_{\odot}$ & $^{4}{\rm He}, ^{12}{\rm C},^{20}{\rm Ne}$ &
0.02 & 0.0 & $2.0\times10^{-2}$& Partially burnt Helium in the He  \\
& & & & &  &core beyond the ext. edge of  C+\\
& & & & &  &O core upto the edge of He core\\
& $(8.1-8.3)$M$_{\odot}$ & $^{4}{\rm He},^{14}{\rm N},^{20}{\rm Ne} $ & 0.0017
& 0.005 & $3.4 \times 10^{-4}$ & Unburnt He-core. Product from\\
& & & & & & CNO H-burning\cr
\enddata
\tablenotetext{a} { Compare with Neon mass obtained from the
 Chandra spectrum $\sim (0.5-1.0)\times 10^{-2} M_\odot$.}
\end{deluxetable}

\clearpage

\clearpage

\clearpage

\clearpage

\clearpage

\clearpage

\clearpage

\clearpage

\clearpage

\clearpage

\begin{thebibliography}{}

\bibitem[Arnaud(1996)]{arn96} Arnaud, K. A. 1996, Astronomical 
Data Analysis Software and Systems V, A.S.P. Conf. series, Vol. 101, 
eds. G. H. Jacoby \& Jeannette Barnes, p. 17
\bibitem[Caughlan \& Fowler(1988)]{cau88} Caughlan, G. R., \& Fowler, W. A.
1988, Atomic Data and Nuclear Data Tables, Vol. 40, 283
\bibitem[Chandra \& Ray(2005)]{cha05} Chandra, P., \& Ray, A. 2005, 
in preparation
\bibitem[Chevalier \& Fransson(2003)]{che03} Chevalier, R., Fransson, C.
2003, Supernovae and Gamma-Ray Bursters, eds. K. Weiler., 
Lecture Notes in Physics, vol. 598, p.171-194
\bibitem[Chugai \& Danziger(2003)]{chu03} Chugai, N. N., Danziger, I. J. 
2003, Astron. Lett. 29, 649 
\bibitem[Chugai(1993)]{chu93} Chugai, N. N. 1993, Astron. Rep., 41, 672
\bibitem[Dickey \& Lockman(1990)]{dic90} Dickey, J. M.\& Lockman, F. J. 1990,
\araa, 28, 215
\bibitem[Fabian \& Terlevich (1996)]{fab96} Fabian, A. C. \& 
 Terlevich, R. 1996, \mnras, 280, L5
\bibitem[Filippenko(1997)]{fil97} Filippenko, A. V. 1997, \araa, 35, 309
\bibitem[Filippenko(1991)]{fil91} Filippenko, A. V. 1991, Supernova 1987A 
and other supernovae, ESO Conference and Workshop Proceedings, eds.  
J. Danziger, and Kurt Kjaer., p.343 
\bibitem[Fox et al.(2000)]{fox00} Fox, D. W., Lewin, W. H. G., Fabian, A.,
et al. 2000, \mnras, 319, 1154
\bibitem[Fransson et al.(2002)]{fra02} Fransson, C., Chevalier, R., 
Filippenko, A. V., et al. 2002, \apj, 572, 350
\bibitem[Fransson, Lundqvist \& Chevalier (1996)]{fra96} Fransson, C., 
Lundqvist,P. \& Chevalier, R., 1996, \apj, 461, 993
\bibitem[Garnavich \& Challis(1995)]{gar95} Garnavich, P., Challis, P.,
\&  Berlind, P. 1995, IAU Circ. 6174
\bibitem[Giesen et al.(1993)]{gie93} Giesen, U., Browne, C.P.,
Gorres, J., Graff, S., et al. 1993, Nucl. Phys. A, 561, 95
\bibitem[Henry \& Branch(1987)]{hen87} Henry, R. B. C., Branch, D. 1987, PASP, 99, 112
\bibitem[Immler \& Lewin(2003)]{imm03} Immler, S., \& Lewin, W. H. G. 2003,
in Lecture Notes in Physics 598, ``Supernovae and Gamma-Ray Bursters", 
p. 91, ed. K. Weiler (springer)
\bibitem[Kaastra(1992)]{kaa92} Kaastra, J.S. 1992, An X-Ray 
Spectral Code for Optically Thin
Plasmas (Internal SRON-Leiden Report, updated version 2.0)
\bibitem[Li et al.(2002)]{li02} Li, W., Filippenko, A. V., Van Dyk, S. D., 
\& Hu, J. 2002, \pasp, 114, 403
Ed: Jan van Paradijs, Johan A. M. Bleeker, Springer
\bibitem[Liedahl et al.(1995)]{lie95} Liedahl, D. A., 
Osterheld, A. L., \& Goldstein, W. H. 1995, \apj, 438, L115
\bibitem[Liedahl et al.(1992)]{lie92} Liedahl, D., 
Kahn, S. M., Osterheld, A. L., \& Goldstein, W. H. 1992,
ApJ, 391, 306
\bibitem[McCray(1987)]{mcc87} McCray, R. A. 1987, in
Spectroscopy of Astrophysical plasmas, ed. A. Dalgarno \&
D. Layzer
\bibitem[Mewe et al.(1985)]{mew85} Mewe, R., Gronenschild, E.H.B.M., \& 
van den Oord, G.H.J. 1985, A\&AS, 62, 197
\bibitem[Mewe et al.(1986)]{mew86} Mewe, R., Lemen, J.R., \& van den 
Oord, G.H.J.  1986, A\&AS, 65, 511
\bibitem[Mewe(1999)]{mew99} Mewe, R. 1999, in X-ray spectroscopy in
Astrophysics, Lecture Notes in Physics,
 Ed: Jan van Paradijs, Johan A. M. Bleeker, Springer
\bibitem[Nomoto et al.(1996)]{nom96} Nomoto, K., Iwamoto, K., Suzuki, T.
et al. 1996 in Compact Stars in Binaries, IAU Symp. 165,
eds. J. van Paradijs et al, Kluwer, Dordrecht
\bibitem[Osterbrock(1989)]{ost89} Ostrebrock, D. E. 1989, Astrophysics of
Gaseous Nebulae and Active Galactic Nuclei, University Science Books, 
Mill Valley
\bibitem[Pollas(1995)]{pol95} Pollas, C., Albanese, D., 
Benetti, S., Bouchet, P. \& Schwarz, H. 1995, IAU Circ. 6170
\bibitem[Pooley \& Lewin(2003)]{poo03} Pooley, D., Lewin, \& W. H. G.
2003, ATel, 116, 1
\bibitem[Pooley et al.(2002)]{poo02} Pooley, D., Lewin, W. H. G., Fox, D. W.,
et al. 2002, \apj, 572, 932
\bibitem[Predehl \& Schmitt(1995)]{pre95} Predehl, P., \& Schmitt, J. H. M. M.
1995, A\&A, 293, 889
\bibitem[Rolfs \& Rodney(1988)]{rol88} Rolfs, C. E., \& Rodney, W. S. 1988,
{\it Cauldrons in the Cosmos}, 
Univ. of Chicago Press.
\bibitem[Schlegel et al.(2004)]{sch04} Schlegel, E. M., Kong, A., Kaaert, P., 
DiStefano, R. \& Murray, S. 2004, \apj,  603, 644
\bibitem[Schlegel(1995)]{sch95} Schlegel, E. M. 1995, Rep. Prog. Phys., 58,
1375
\bibitem[Van Dyk et al.(1996)]{van96} Van Dyk, S. D., Weiler, K. W., 
Sramek, R. A., et al. 1996, \aj, 111, 1271
\bibitem[Verner \& Ferland(1996)]{ver96} Verner, D. A. \& Ferland, G. J.
1996, \apjs, 103, 467
\bibitem[Woosley et al.(2002)]{woo02} Woosley, S. E., Heger, A., Weaver, T. A.
2002, Rev. Mod. Phys. 74, 1015
\end{thebibliography}
\end{document}